\documentclass[%
reprint, superscriptaddress, amsmath, amssymb, aps]{revtex4-2}

\usepackage{graphicx}
\usepackage{dcolumn}
\usepackage{bm}
\usepackage{hyperref} \hypersetup{ colorlinks=true, linkcolor=blue, filecolor=blue, urlcolor=blue, citecolor = blue, pdfpagemode=FullScreen}

\begin{document}

\preprint{APS/123-QED}

\title{A superconducting quantum information processor with high qubit connectivity}

\author{Gürkan Kartal}\altaffiliation[Alumnus of ]{Koç University}


\email{gkartal@ku.edu.tr}%
\author{George Simion}%
\affiliation{%
IMEC, 3001 Leuven, Belgium
}%

\author{Bart Sorée}
\affiliation{%
	IMEC, 3001 Leuven, Belgium
}%
\affiliation{
Department of Physics, University of Antwerp, 2020 Antwerp, Belgium
}%
\affiliation{
Electrical Engineering Department, KU Leuven, 3001 Leuven, Belgium
}%

\date{\today}

\begin{abstract}
Coupling of transmon qubits to resonators that serve as storage for information provides alternative routes for quantum computing. Such a scheme paves the way for achieving high qubit connectivity, which is a great challenge in cQED systems. Implementations either involve an ancillary transmon’s direct excitation, or virtual photon interactions. Virtual coupling scheme promises advantages such as the parallel, virtual gate operations and better coherence properties since the transmon’s decoherence effects are suppressed. However, virtual gates rely on nonuniform frequency separation of the modes in the system and acquiring this feature is not a straightforward task. Here, we propose an architecture that incorporates the four-wave mixing capabilities of the transmon into a chain of resonators coupled collectively by qubits in between. The system, consisting of numerous resonators all operating within the single mode approximation, maintains the above-mentioned nonuniformity by accommodating different resonators with appropriate frequencies. 

\end{abstract}

\maketitle

\section{\label{sec:level1}Introduction}
Circuit QED (cQED) provides the means to realize quantum computing architectures that are among today's most prominent ones \cite{Blais2007, Schoelkopf2008}. In this medium, coupling of superconducting qubits to resonators that serve as quantum memories leads to alternative routes for quantum computing: nonlinear superconducting elements allow for the initialization, manipulation, readout as well as protection of quantum states preserved in such memories \cite{Reagor2016, Hofheinz2008, Krastanov2015, Heeres2017, Sun2014, Ofek2016, Hu2019}. Furthermore, various architectures with superconducting resonator arrays (which can potentially enhance the capabilities of previous designs) have been reported previously \cite{Sun2006, Mariantoni2008, Reuther2010, Baust2015, Peropadre2013, Wulschner2016, Raussendorf2007,  Helmer_2009, DiVincenzo_2009,  Johnson2010, Koch2010, Steffen2011, Wang2011, Mariantoni2011, Underwood2012}. It has been also demonstrated that multiple qubits and resonators can be fabricated in the same circuit \cite{Fink2009, Filipp2011, vanLoo2013, Mlynek2014, Lambert2016}.

\begin{figure}
	\centering
	\includegraphics[scale=0.4]{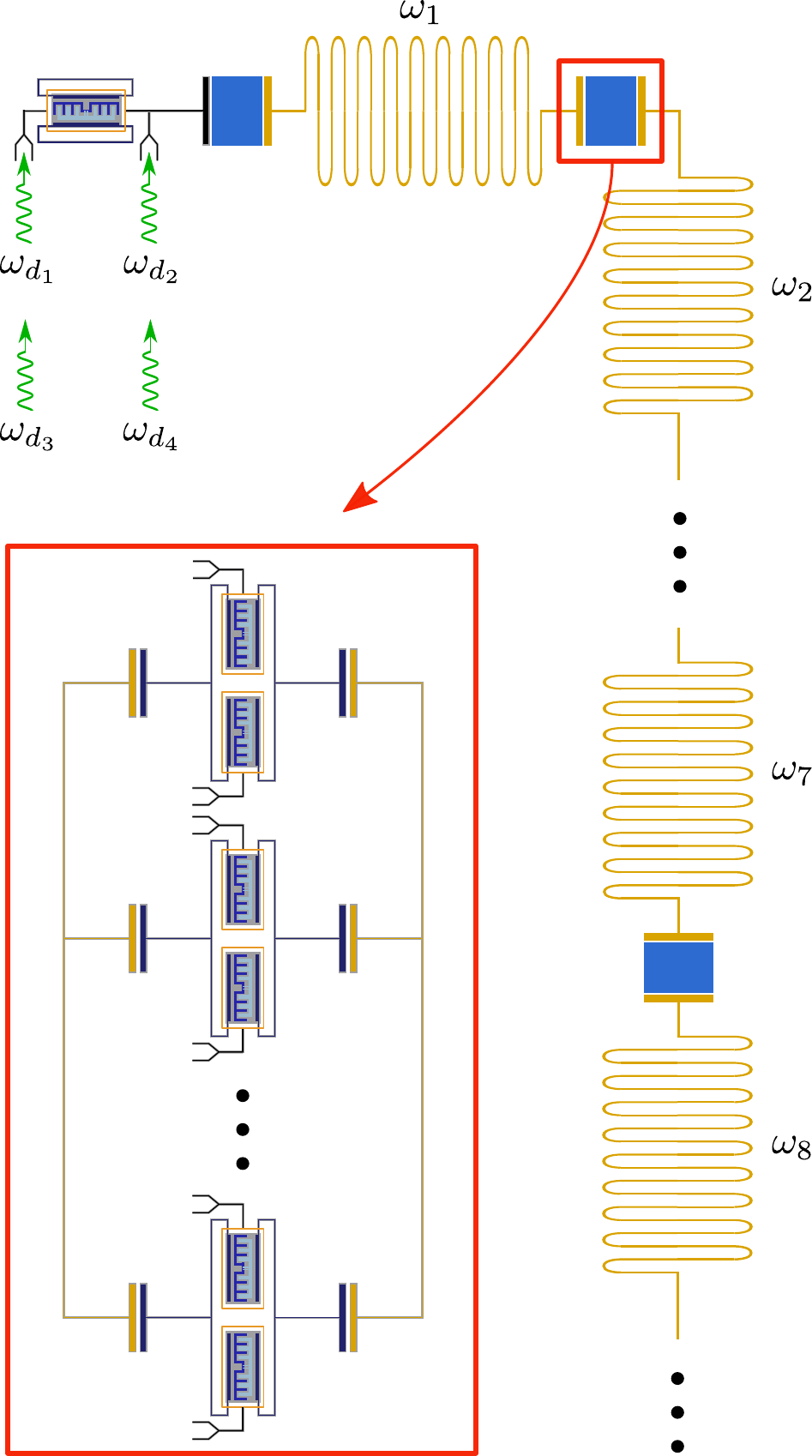}
	\caption{Schematic of the quantum information processor. Superconducting resonators (yellow) are capacitively coupled through multiple transmon qubit pairs (blue). Quantum gates are implemented by applying microwave drives (green) to an ancillary transmon coupled to the first resonator in the chain through another group of transmon pairs. Nonuniform frequency separation of the modes (required for the quantum operations) is maintained by accommodating resonators with appropriate frequencies.}
	\label{fig:Gurkan}
\end{figure}

\begin{figure}
	\centering
	\includegraphics[scale=0.4]{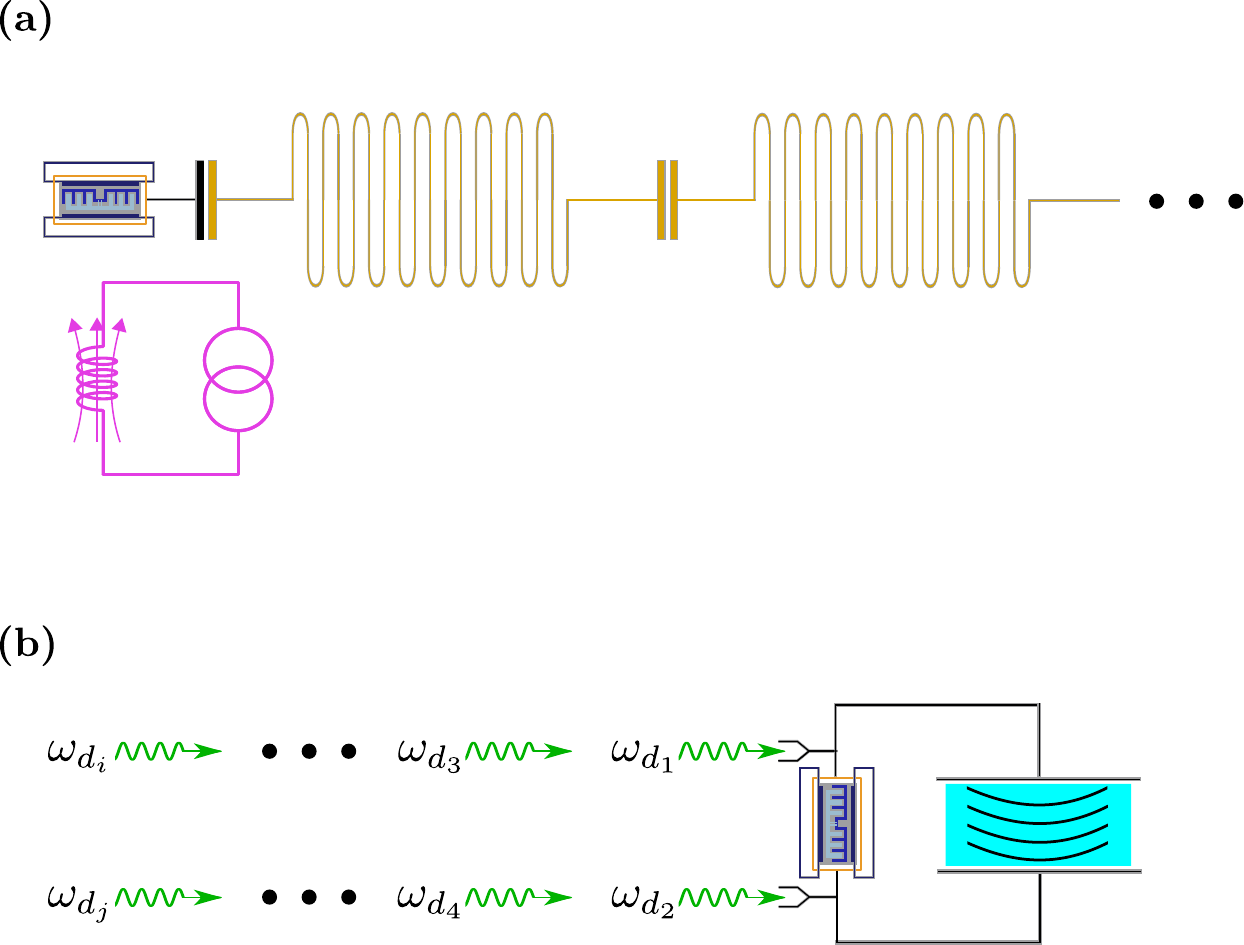}
	\caption{Previously proposed solutions for the qubit connectivity issue. (a) A quantum processor, in which an ancillary transmon's frequency is modulated to implement quantum gates between the qubits encoded in superconducting resonator modes. These operations involve direct excitation of the transmon, which is disadvantageous in terms of decoherence properties, and also prevents parallel gate operations \cite{Naik2017}. (b) An example of a cQAD quantum processor consisting of an ancillary transmon and a bulk acoustic wave resonator. This setup allows for implementing a universal gate set for the qubits encoded in acoustic modes exploiting tranmon's four-wave mixing capabilities. During the quantum operations, the transmon is only virtually excited, which leads to mitigation of the decoherence effects (dominated by the transmon) and implementation of gates in parallel. These operations require nonuniform frequency separation of the acoustic modes, which needs to be engineered \cite{Hann2019}. The system Hamiltonian (see Eq. \ref{equation:Gurkan}) is in the same form as the effective Hamiltonian of the structure given in Fig. \ref{fig:Gurkan}.}
	\label{fig:Why}
\end{figure}

Quantum information processing in superconducting circuits requires tunable coupling between various components, i.e., qubits, or a qubit and a resonator, or resonators \cite{Sun2006, Mariantoni2008, Reuther2010, Baust2015, Peropadre2013, Wulschner2016, Blais2003, Berkley2003, McDermott2005, Liu2006, Hime2006, vanderPloeg2007, Majer2007, Bialczak2011, Cleland2004, Sillanpaa2007, Leek2010, Eichler2012, Allman2014, Lu2017, Yin2013}. Such couplings in the state-of-the-art architectures occur through nearest-neighbor interactions. This limits the qubit connectivity leading to overhead in computations. We make a proposal (Fig. \ref{fig:Gurkan}) to address the connectivity issue, and our work is related to existing ones, which either involve an ancillary transmon’s direct excitation \cite{Naik2017} (Fig. \ref{fig:Why} (a)), or virtual photon interactions \cite{Hann2019} (Fig. \ref{fig:Why} (b)). The former, a multimode cQED approach, involves swapping the information into the transmon (and exciting it) to implement quantum gates between qubits encoded in resonator modes. The latter proposes a multimode system in circuit quantum acoustodynamics (cQAD) \cite{O’Connell2010, Pirkkalainen2013, Gustafsson2014, Chu2017, Chu2018, Kervinen2018, Manenti2017, Noguchi2017, Satzinger2018, Moores2018, Bolgar2018, Sletten2019, Arrangoiz-Arriola2019} (whose methodology is also applicable to multimode cQED systems \cite{Hann2019}), which involves applying off-resonant drives to an ancillary transmon initiating interactions (or quantum operations) between the qubits encoded in acoustic modes. Here, the ancillary transmon acts as a four-wave mixer due to its Kerr nonlinearity \cite{Mutus2013, Leghtas2015, Gao2018, Zhang2019}, converting phonons from one frequency to another. During these interactions, which allow for implementing a universal gate set for these so called phononic qubits, the transmon is only virtually populated. Compared to the former, this virtual coupling scheme promises advantages such as (i) parallel gate operations; and (ii) substantial improvement in gate fidelities since the transmon's decoherence is mitigated. These virtual gates require nonuniform frequency separation of the modes in the system ensuring that coupling of the modes can be initiated selectively, i.e., resonance conditions are not degenerate. However, acquiring this nonuniformity is not a straightforward task and obtained through methods such as creating (a) point defect, or periodic nonuniformities (for the case of bulk or surface acoustic wave resonators); or (b) engineering the geometry of the structure (for the case of phononic crystal resonators) \cite{Hann2019}.

In this work, we suggest an alternative approach and propose a superconducting quantum information processor architecture that exploits the four-wave mixing capabilities of an ancillary transmon connected to a chain of resonators. These resonators, all operating within the single mode approximation, are coupled by a collection of two-level atoms (transmons) in between. The collection of two-level atoms increase the strength of the coupling between the ancillary transmon and various (non-neighboring) resonators. By applying off-resonant drives to the ancillary transmon, interactions between the resonator modes are initiated in the same way as in the system of Ref. \cite{Hann2019}. These interactions are then used to implement a universal quantum gate set. The above-mentioned nonuniformity in the system is maintained by accommodating different resonators with appropriate frequencies.

\section{Theoretical Description of the Quantum Processor}

The system, as schematically shown in Fig. \ref{fig:Gurkan}, consists of an ancillary transmon along with a chain of resonators that are collectively coupled to each other and the ancillary transmon through two-level artificial atoms, e.g., transmon pairs. Our work was inspired from (i) Ref. \cite{Chen2018}, which proposes a system in which coupling between non-neighboring superconducting resonators (which belong to a chain of resonators) can be initiated. There, the resonators in the chain are coupled to each other through two-level artificial atoms, which together behave as a multilevel atom \cite{Dicke1954, Gross1982, Fink2009, Filipp2011, vanLoo2013, Freedhoff1967, Stroud1972, Pavolini1985}; and (ii) Ref. \cite{Hann2019} (see Fig. \ref{fig:Why} (b)), where an ancillary transmon is coupled to an acoustic resonator and the quantum operations between the qubits encoded in acoustic modes are performed by applying off-resonant drives to the transmon. The system can be described by the Hamiltonian
\begin{eqnarray}
	H=&&\omega_{q} q^{\dagger} q - \frac{\alpha}{2}(q^{\dagger} q)^2  \nonumber\\
	&&+ \sum_{k} [\omega_{k} m^{\dagger}_{k}m_{k} + {g}_{k} (q^\dagger m_{k} + q m_{k}^\dagger)] + H_{d}.
	\label{equation:Gurkan}
\end{eqnarray}
Here, $H_{d} = \sum_{j} \Omega_{j} q^{\dagger} \exp(-i \omega_{j}t) +$H.c. denotes external drives on the transmon, whereas $q$ and $m_{k}$ are annihilation and creation operators for the transmon and acoustic modes, respectively. The transmon, having Kerr nonlinearity $\alpha$, is coupled to the $k$th phonon mode with strength $g_{k}$. Main result of our work will be showing that the effective Hamiltonian of the system in Fig. \ref{fig:Gurkan} is in the same form of Eq. \ref{equation:Gurkan}. There, (a) $q$ will depict annihilation and creation operators for the modes of the superconducting resonators operating within the single mode approximation; (b) transmon's couplings to these modes with strengths $g_{k}$ will be magnified proportional to the number of artificial atoms (transmons) between the resonators.

\begin{figure}
	\centering
	\includegraphics[scale=0.058]{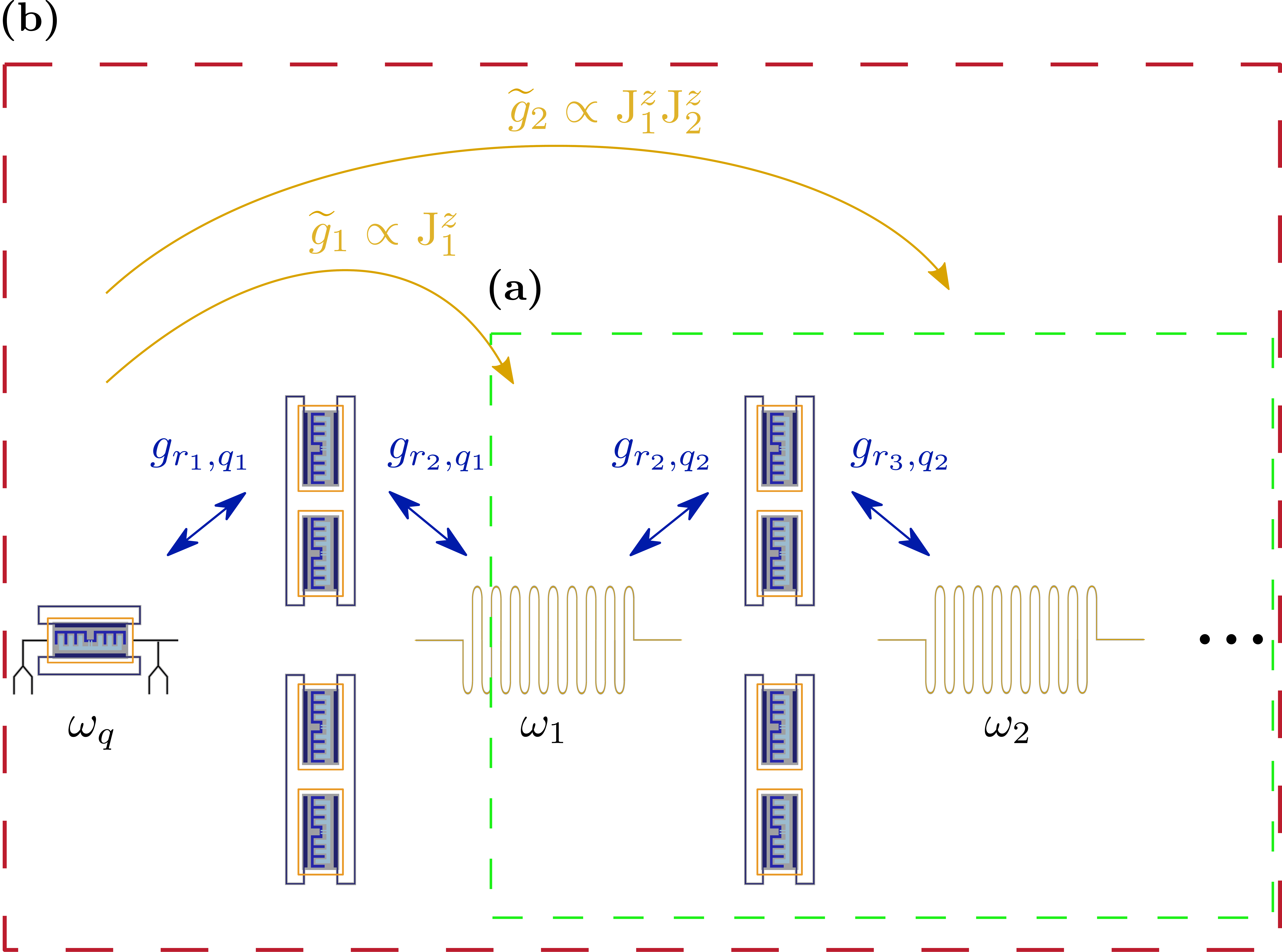}
	\caption{Conceptual diagram illustrating coupling strengths between various components of the system. (a) A resonator's coupling strength to another can be amplified by coupling these resonators collectively by transmons in between. This is also true for non-neighboring resonators residing in a chain-like arrangement (see Section \ref{subsec:3A}) \cite{Chen2018}. (b) Ancilla transmon's coupling strength to a resonator residing in a chain-like arrangement can be magnified by collective qubit coupling. Effective Hamiltonian of the system is in the same form as the one given in Eq. \ref{equation:Gurkan}, and the coupling strengths are proportional to the number of transmons used as couplers, i.e., $\widetilde{g}_{1} \propto $ J$_{1}^{z}$, $\widetilde{g}_{2} \propto $ J$_{1}^{z}$J$_{2}^{z}$,..., $\widetilde{g}_{k} \propto \prod_{i=1}^{k} $ J$_{i}^{z}$ (see Section \ref{subsec:3B}).}
	\label{fig:Result}
\end{figure}
Such a system with the Hamiltonian given in Eq. \ref{equation:Gurkan} allows to initialize, manipulate and measure qubits encoded in resonator modes \cite{Pechal2014, Srinivasan2014, Axline2018, Kurpiers2018, Chu2018, Satzinger2018}. Furthermore, applying off-resonant drives to ancillary transmon (of the system in Fig. \ref{fig:Why} (b), for example) initiates interactions (or quantum operations) between these qubits. These interactions are realized due to the Kerr nonlinearity of the transmon, which acts as a four-wave mixer converting phonons (or microwave photons) from one frequency to another. Adopting a linear optical quantum computing approach, i.e., dual-rail encoding scheme \cite{Knill2001}, two types of such interactions are needed for computations. First one occurs, for example, by applying two drives with frequencies $\omega_{d_{1},d_{2}}$ satisfying the resonance condition $\omega_{d_{2}} - \omega_{d_{1}} = \omega_{2} - \omega_{1} $, which leads to the effective Hamiltonian $H = g^{(1)}_{v} m_{1}m^{\dagger}_{2} + $ H.c., where $g^{(1)}_{v}$ is the virtual coupling strength. This two-mode interaction can be used to implement arbitrary rotation of single qubits encoded in $|0,1\rangle$ Fock states. Second one is the result of a single drive with frequency $\omega_{d_{1}} = \omega_{2} + \omega_{8} - \omega_{3}$ leading to the Hamiltonian $H = g^{(2)}_{v} m_{2}m_{8}m^{\dagger}_{3} + $ H.c., where $g^{(2)}_{v}$ is the virtual coupling strength. Considering two qubits encoded in four modes, i.e. $\{1,2\}$ and $\{7,8\}$, respectively, along with an ancilla mode $3$, this three-mode interaction can be used to implement a controlled phase (CZ) gate between these qubits. Assuming $|\bar{0}\rangle = |1\rangle_{1}|0\rangle_{2}$ and $|\bar{1}\rangle = |0\rangle_{1}|1\rangle_{2}$ are the states of the single qubit encoded in modes $1$ and $2$, and $|\bar{0}\bar{1}\rangle$ is a two-qubit state (corresponding, for example, to two qubits encoded in modes $1, 2, 7, 8$); evolution under the latter Hamiltonian for a time $\pi/g^{(2)}_{v}$ imparts a $-1$ phase to the state $|\bar{1}\bar{1}\rangle$ implementing a CZ gate. These two operations, i.e., arbitrary single-qubit rotations and CZ form a universal gate set \cite{Niu2018}. Additionally, multiple of these gates can be implemented in parallel by applying multiple microwave drives, as seen in Figs. \ref{fig:Gurkan} and \ref{fig:Why} (b) \cite{Hann2019}.    

\section{Assembling the System}

We will first discuss the collective (transmon) qubit coupling of resonators (see Fig. \ref{fig:Result} (a)-green dashed rectangle), as well as provide a numerical example in Subsection \ref{subsec:3A}. Subsequently, we will present the proposed system (see Fig. \ref{fig:Result} (b)-red dashed rectangle), in which the four-wave mixing capabilities of the transmon is incorporated into a chain of resonators coupled collectively by qubits in between in Subsection \ref{subsec:3B}.

\subsection{Collective Qubit Coupling of Resonators}
\label{subsec:3A}

We consider the system of a resonator chain, whose coupling scheme is given in Fig. \ref{fig:Result} (a). Supposing that three resonators with frequencies $\omega_{1}, \omega_{2}$, and $\omega_{3}$ are present in the structure, where $\omega_{1} = \omega_{3} $, and the coupler transmons are prepared in the pairwise subradiant states, the effective Hamiltonian of the system reads \cite{Chen2018} 
\begin{eqnarray}
	H \approx && \sum_{k=1}^{2} \bigg\{\omega_{k}m_{k}^{\dagger}m_{k} + \frac{1}{2}\bigg[\omega_{q_{k+1}}+2\bigg(\chi_{r_{k+1}, q_{k+1}}\nonumber\\
	&&+\frac{(g_{r_{k}, r_{k+1}})^{2}}{\Delta_{r_{k}, r_{k+1}}}\textbf{J$_{k}^{z}$}\bigg)m_{k}^{\dagger}m_{k} +2\bigg(\chi_{r_{k+2}, q_{k+1}} \nonumber\\
	&&-\frac{(g_{r_{k}, r_{k+1}})^{2}}{\Delta_{r_{k}, r_{k+1}}}\textbf{J$_{k}^{z}$}\bigg)m_{k+1}^{\dagger}m_{k+1}\bigg] \textbf{J$_{k}^{z}$}\bigg \} \nonumber\\
	&& + \frac{g_{r_{1}, r_{2}}g_{r_{2}, r_{3}}}{\Delta_{r_{k}, r_{k+1}}} (m_{1}^\dag m_{3} + m_{3} m_{1}^\dag)\textbf{J$_{1}^{z}$}\textbf{J$_{2}^{z}$},\nonumber\\
	\label{equation:resonator_chain}
\end{eqnarray}
where 
\begin{eqnarray}
	\chi_{r_{i}, q_{j}} = && (g_{r_{i}, q_{j}})^2/\Delta_{r_{i}, q_{j}},\nonumber\\
	g_{r_{i}, r_{i+1}} = && \frac{\Delta_{r_{i}, q_{i}}+\Delta_{r_{i}, r_{i+1}}}{2\Delta_{r_{i}, r_{i+1}}\Delta_{r_{i}, r_{i+1}}},\nonumber\\
	\textbf{J$_{i}^{z}$} = && \sum_{j=1}^{N/2} J^{z}_{j}.\nonumber
\end{eqnarray}

\begin{figure}
	\centering
	\includegraphics[scale=0.4]{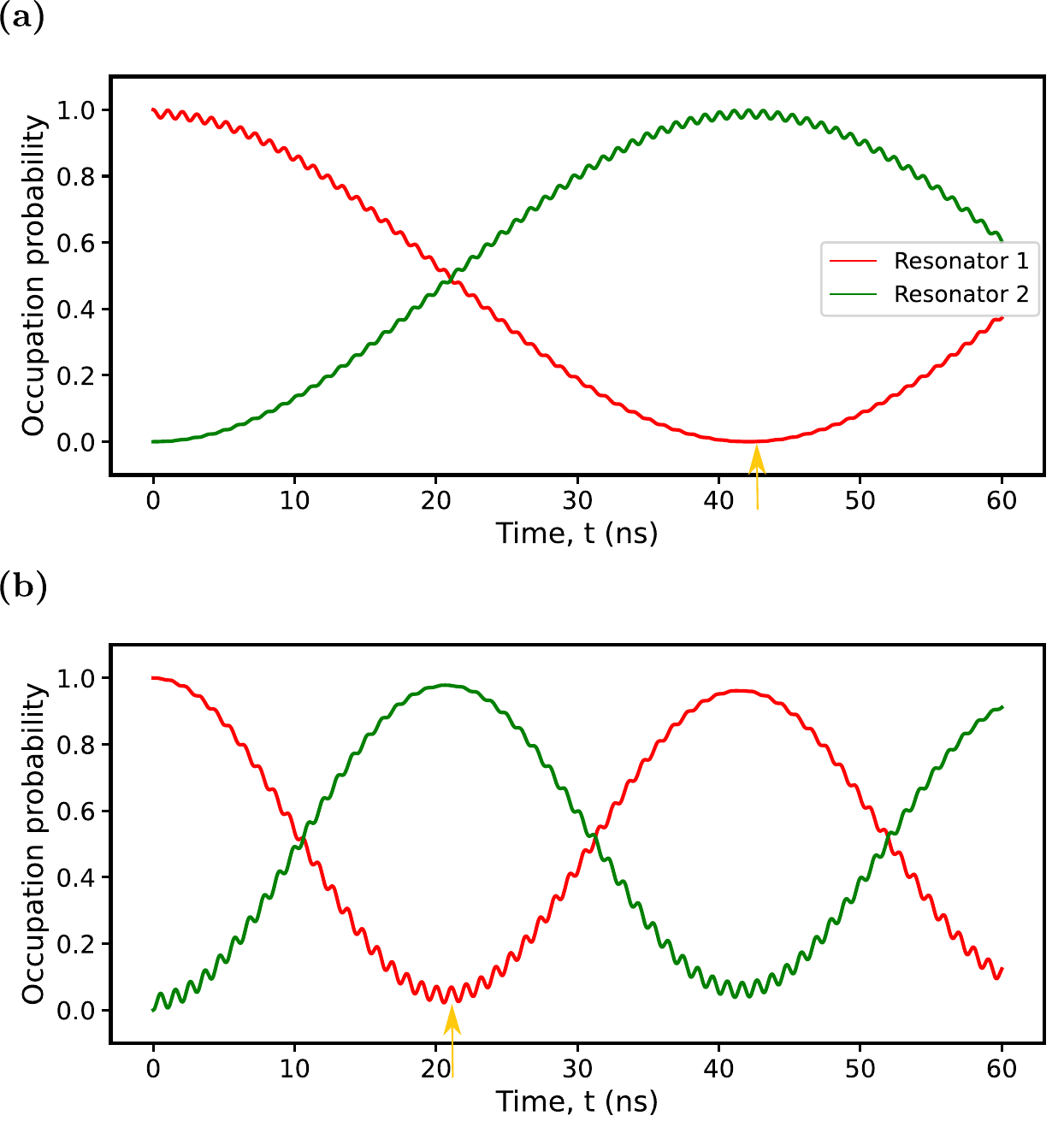}
	\caption{The master equation simulation of the system presented in Fig. \ref{fig:Result} (a), for two cases where two resonators with frequencies $\omega_{1}$ and $\omega_{2}$ are coupled by a (a) single transmon; and (b) pair of transmons, respectively. A photon swap interaction takes place with the duration of $t=\pi / 2 \tilde{g}$, where $\tilde{g}$ is the effective coupling strength between resonators. Going from one transmon coupler to two, $\tilde{g}$ doubles, hence, $t$ goes from 43.75 ns to 21.88 ns, as indicated by the yellow arrows. The system parameters are chosen as follows: $\omega_{1,2}/ 2\pi = 4$ GHz, $\omega_{q_{2}}/ 2\pi = 5$ GHz, $\kappa_{1,3}/2\pi = 10$ KHz, $\gamma_{q_{2}}/ 2\pi = 1$ MHz. The capacitance values of the resonators and transmon(s) are taken as $C_{1,2}= 70$ fF and $C_{q_{2}}=200$ fF, respectively; whereas resonator-transmon coupling capacitance values are $C_{r_{2},q_{2}}=C_{r_{3},q_{2}} = 4$ fF.}
	\label{fig:rchain}
\end{figure}

Here, $g_{r_{i}, q_{j}}$ is the strength of the coupling of i$^{th}$ resonator to the j$^{th}$ (coupler) transmon group (except for $g_{r_{1}, q_{1}}$, which denotes the strength of the coupling of ancilla transmon to the first transmon group, see Subsection \ref{subsec:3B}). For simplicity, coupler transmons in the same group are assumed to have the same frequency, denoted by $\omega_{q_{k}}$, and $\Delta_{r_{i}, q_{j}}= \omega_{i} - \omega_{q_{j}} $ (except for $\Delta_{r_{1}, q_{1}}$, which is equal to $\omega_{q} - \omega_{q_{1}}$), and $\Delta_{r_{i}, r_{i+1}}$ denotes the frequency differences of the successive resonators. The operator $\textbf{J$_{i}^{z}$}$ denotes the collective angular momentum of the i$^{th}$ transmon group. In such a group consisting of $N$ transmon pairs, the operator $J^{z}_{j} = (\sigma^{z}_{2k-1} + \sigma^{z}_{2k})$ corresponds to the j$^{th}$ transmon pair. For example, suppose there are 3 transmon pairs in the first group of couplers. Considering that eigenvalues of the operator $\sigma$ are $1$ and $-1$, leaving all of these transmons in the ground state leads to $\textbf{J$_{1}^{z}$}$ being equal to 6; on the other hand, exciting one of the each transmon pairs results in $\textbf{J$_{1}^{z}$}$ being equal to 0. This mechanism can be used to enhance the coupling strength between (non-neighboring) elements in the circuit, as well as a switch isolating certain parts of the structure, and providing the system with a programmability feature.

\begin{figure}
	\centering
	\includegraphics[scale=0.4]{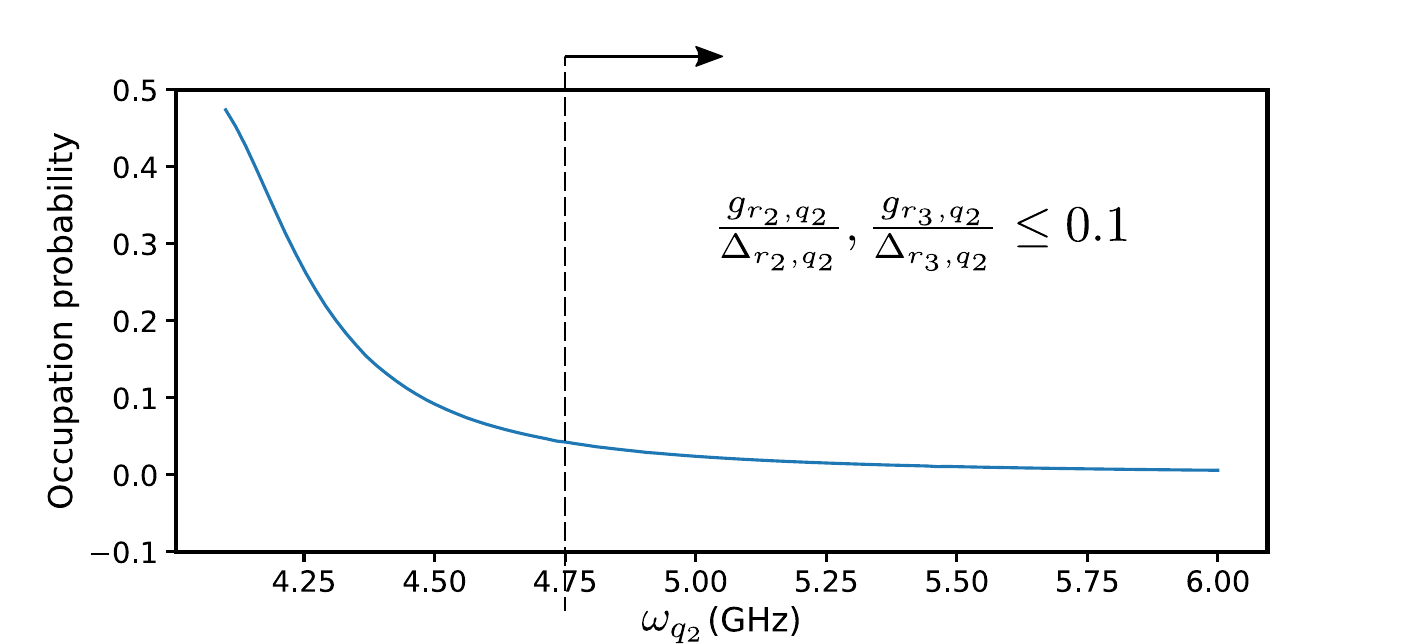}
	\caption{Transmon’s excitation in the system presented in Fig. \ref{fig:Result} (a), for the case where two resonators with frequencies $\omega_{1}$ and $\omega_{2}$ are coupled by a single transmon,  whose frequency is varied. A successive set of master equation simulations is performed for various transmon frequencies	between 4.1 and 6 GHz. Passing of the 4.75 GHz threshold, which is dictated by the dispersive regime requirements, i.e., $g_{r_{2},q_{2}}/\Delta_{r_{2},r_{2}} \leq 0.1$ and $g_{r_{3},q_{2}}/\Delta_{r_{3},r_{2}} \leq 0.1$, leads to a drastic rise in the occupation probability of the transmon ($g_{r_{2},q_{2}}$ and $g_{r_{3},q_{2}}$ varies between 65 and 85 MHz, or on average are equal to 75 MHz).}
	\label{fig:maxcik}
\end{figure}

A photon swap interaction between the first and third resonators can be initiated in this arrangement and generalizing to coupling of the first and n$^{th}$ resonators, the effective interaction in the system can be written as
\begin{eqnarray}\label{equation:interaction}
	V_{r_{1},...,r_{n}} = && g_{r_{1}, r_{n}}(m_{1}^\dag m_{n} + m_{n}^\dag m_{1}) \prod_{i = 1}^{n-1}\textbf{J$_{i}^{z}$},
\end{eqnarray}
where
\begin{eqnarray}
	g_{r_{1}, r_{n}} = && \frac{\prod_{i = 1}^{n-1}g_{r_{i}, r_{i+1}}}{(\Delta_{r_{i}, r_{i+1}})^{n-2}}.
\end{eqnarray}
Suppose that there are $N$ transmon qubits in each coupler group, the strength of the coupling between the first and the n$^{th}$ resonators can be engineered from $0$ to $-N^{n-1}g_{r_{1}, r_{n}}$. As long as $N \geq \Delta_{r_{i}, r_{i+1}}/g_{r_{i}, r_{i+1}}$, strength of the coupling between non-neighboring resonators can be comparable to the coupling strength of neighboring ones \cite{Chen2018}.

The master equation simulation of the system seen in Fig. \ref{fig:Result} (a), for two cases where two resonators with frequencies $\omega_{1}$ and $\omega_{2}$ are coupled by a single transmon and pair of transmons, is presented in Fig. \ref{fig:rchain} (a) and (b), respectively. In both cases, resonator 1 and 2 are initially prepared in Fock states $|1\rangle$ and $|0\rangle$, respectively; and the transmons are in their ground states. A photon swap interaction occurs with the duration of $t=\pi/2\tilde{g}$, where $\tilde{g}$ is the effective coupling strength between the resonators, which doubles as the number of coupler transmons doubles. Consequently, the interaction time, $t$, becomes half of its initial value. If one of the coupler transmons is excited, the interaction is terminated, since $\textbf{J$_{2}^{z}$}$ becomes equal to 0, indicating the \textit{switch} feature of the system.

For the system, where resonators 1 and 2 are coupled by a single transmon, we performed a successive set of master equation simulations for various transmon frequencies, to indicate the amount of unwanted transmon excitation, as seen in Fig. \ref{fig:maxcik}. The transmon's state is vital in the system, since a rise in the occupation probability of the transmon induces \textit{inverse Purcell effect}, where the resonator decay rates are enhanced due to strong coupling to the transmon \cite{Reagor2016}. As expected from the dispersive regime requirements, going beyond the conditions $g_{r_{2},q_{2}}/\Delta_{r_{2},r_{2}} \leq 0.1$ and $g_{r_{3},q_{2}}/\Delta_{r_{3},r_{2}} \leq 0.1$ leads to a drastic rise in the transmon's occupation probability. These conditions will add to the design considerations of the proposed system discussed in Subsection \ref{subsec:3B}.
  
\subsection{Proposed Design}
\label{subsec:3B}

The system in the previous subsection (see Fig. \ref{fig:Result} (a)) provides a mechanism that allows to couple non-neighboring (same frequency) resonators by enhancing their coupling strength via two-level artificial atoms in between. The idea of the proposed design is realizing a system with a Hamiltonian that is in the same form as the one given in Eq. \ref{equation:Gurkan}, where the strength of ancilla transmon's couplings to the individual resonators are enhanced by the same collective qubit coupling mechanism.

Suppose that we modify the previous system by attaching an ancilla transmon to the first resonator in the chain, as seen in Fig. \ref{fig:Result} (b). Assuming there are three resonators with different frequencies present in the system, the dispersive Hamiltonian of the system can be derived in a similar manner to the derivations in Ref. \cite{Chen2018}, and it reads
\begin{eqnarray}
	H \approx &&\omega_{q}^{'}  m_{0}^\dag m_{0} - \frac{\alpha }{2} (m_{0}^\dag m_{0})^2 + \sum_{k=1}^{3}\bigg[\omega_{k}m_{k}^\dag m_{k}+\frac{1}{2}(\omega_{q_{k}}   \nonumber\\
	&&+ 2 \chi_{r_{k}, q_{k}} m_{k-1}^\dag m_{k-1} + 2 \chi_{r_{k+1}, q_{k}} m_{k}^\dag m_{k})\textbf{J$_{k}^{z}$}\nonumber\\
	&& + g_{r_{k},r_{k+1}} (m_{k-1}^\dag m_{k} + m_{k-1} m_{k}^\dag)\textbf{J$_{k}^{z}$}\bigg],
	\label{equation:Gurkan_final}
\end{eqnarray}
where $\omega_{q}^{'} = \omega_{q} - \alpha/2 $, and $m_{0}$ corresponds to the operator $q$. To obtain an Hamiltonian where the interaction terms corresponding to the couplings between resonators are effectively eliminated, we apply the following unitary transformations on Eq. \ref{equation:Gurkan_final}, consecutively:
\begin{align}\label{eq:unit1}
	U_{1\rightarrow 2} &=\text{exp}\bigg[\frac{g_{r_{2},r_{3}} ( {m_{2}} m_{1}^\dag-m_{2}^\dag {m_{1}} )\textbf{J$_{2}^{z}$}}{\Delta_{r_{2},r_{3}}}\bigg],
\end{align}
and 
\begin{align}
	U_{2\rightarrow 3} 
	=&
	\text{ exp}\bigg[\frac{g_{r_{2},r_{3}} g_{r_{3},r_{4}}}{\Delta_{r_{2},r_{3}} (\Delta_{r_{2},r_{3}}+\Delta_{r_{3},r_{4}})}\nonumber\\
	& \times (- m_{3}^\dag m_{1} + m_{3} m_{1}^\dag)\textbf{J$_{2}^{z}$}\textbf{J$_{3}^{z}$} \nonumber\\
	& + \frac{g_{r_{3},r_{4}}}{\Delta_{r_{3},r_{4}}}(- m_{3}^\dag m_{2} + m_{3} m_{2}^\dag)\textbf{J$_{3}^{z}$}\bigg].
\end{align}
Using Baker-Campbell-Hausdorff formula with a second order expansion, we obtain an effective Hamiltonian which reads 
\begin{equation}\label{eq:mainresult}
	\hat{H}=\widetilde{\omega}_{q} q^{\dagger} q - \frac{\alpha}{2} (q^{\dagger} q)^2  + \sum_{k=1}^{3}\big[\widetilde{\omega}_{k} m_{k}^{\dagger} m_{k}+ \widetilde{g}_{k} (q^\dagger m_{k} + q m_{k}^\dagger)\big],
\end{equation}
where
\begin{align}\label{eq:inas}
	&\widetilde{\omega}_{q}= \omega_{q} - \frac{3\alpha}{2} ,\\
	&\widetilde{\omega}_{1}= \omega_{1} +\frac{g_{r_{2},r_{3}}^{2}}{\Delta_{r_{2},r_{3}}}(\textbf{J$_{2}^{z}$})^{2} ,\\
	&\widetilde{\omega}_{2}= \omega_{2} -\frac{g_{r_{2},r_{3}}^{2}}{\Delta_{r_{2},r_{3}}}(\textbf{J$_{2}^{z}$})^{2} +\frac{g_{r_{3},r_{4}}^{2}}{\Delta_{r_{3},r_{4}}}(\textbf{J$_{3}^{z}$})^{2} + \chi_{r_{3},q_{3}}\textbf{J$_{3}^{z}$},\\
	&\widetilde{\omega}_{3}= \omega_{3}-\frac{g_{r_{2},r_{3}}^{2}}{\Delta_{r_{2},r_{3}}}(\textbf{J$_{3}^{z}$})^{2}+ \chi_{r_{4},q_{3}}\textbf{J$_{3}^{z}$},\\
	&\widetilde{g}_{1} =g_{r_{1},r_{2}}\textbf{J$_{1}^{z}$},\\
	&\widetilde{g}_{2} = - \frac{g_{r_{1},r_{2}}g_{r_{2},r_{3}} }{\Delta_{r_{1},r_{2}}}\textbf{J$_{1}^{z}$} \textbf{J$_{2}^{z}$},\\
	&\widetilde{g}_{3} = \frac{g_{r_{1},r_{2}}g_{r_{2},r_{3}}g_{r_{3},r_{4}}}{\Delta_{r_{2},r_{3}} (\Delta_{r_{1},r_{2}} + \Delta_{r_{2},r_{3}})}\textbf{J$_{1}^{z}$} \textbf{J$_{2}^{z}$}\textbf{J$_{3}^{z}$}.
\end{align}
We can generalize the procedure to $n$ resonators, leading to an interaction term similar to the one given in Eq. \ref{equation:interaction}.

The effective Hamiltonian given in Eq. \ref{eq:mainresult}, which is in the same form as Eq. \ref{equation:Gurkan}, is the main result of this paper. Consequently, the system given in Fig. \ref{fig:Gurkan} allows to initialize, manipulate, and measure qubits encoded in resonator modes, as well as implement operations between such qubits that form a universal quantum gate set.

\section{Conclusions and Discussions}
We have proposed an architecture for quantum computing with multimode cQED systems having high qubit connectivity. The qubits are encoded in resonator modes in accordance with linear optical quantum computing formalism. The resonators operate within the single mode approximation, and quantum operations are performed by applying off-resonant drives to an ancilla (processor) transmon. Such operations require nonuniform frequency separation of the modes in the system, which is maintained by accommodating different resonators with appropriate frequencies, unlike the previous work \cite{Hann2019}, where nonuniformity  has to be engineered. The strength of the ancilla transmon's coupling to various resonators are enhanced by the collective qubit coupling mechanism, which may provide the system with a programmability feature, since the coupler transmon groups can act as a switch (which can be turned off within nanoseconds \cite{Chen2018}) between resonators. This collective qubit coupling feature may also be incorporated into the previous proposal with multimode cQAD systems.

For the system to be viable, certain design parameters arise originating from the unitary transformations to arrive at dispersive Hamiltonians, Eq. \ref{equation:Gurkan_final} and  Eq. \ref{eq:mainresult}. To obtain these two equations, interaction terms corresponding to (i) resonator-transmon group couplings; and (ii) resonator-resonator couplings are effectively eliminated, respectively. These are done with the assumptions of (a) transmons being largely detuned from resonators that are nearest neighbors to them (see Fig. \ref{fig:maxcik}); and (b) resonators in question being detuned from each other. Therefore, the values of parameters such as $g_{r_{1}, q_{1}}/\Delta_{r_{1}, q_{1}}$ and $g_{r_{2},r_{3}}/\Delta_{r_{2},r_{3}}$ should be lower than or equal to 0.1, as suggested in Ref. \cite{Chen2018}.

Another design consideration comes from hybridization of ancilla transmon with the resonator modes, shifting their frequencies. These shifts arise from the drives on the transmon (AC Stark shifts), or Kerr interactions. In the presence of such shifts, nonuniformity in the mode frequency separation needs to be maintained for selective mode coupling, and these issues are addressed in Ref. \cite{Hann2019}. Furthermore, there is drive-induced heating of the ancilla transmon, contributing to the infidelity of operations. Even though, for small drives, infidelity associated with this effect is negligible, stacking up more drives on the transmon to apply more parallel gates, could harm operation fidelity \cite{Hann2019,Zhang2019}. One potential solution for such a problem could be attaching additional ancilla (processor) transmons to the proposed system, replacing some of the resonators that are used as quantum memories. 

\begin{acknowledgments}
	We thank the members of Quantum Computing Group at IMEC for helpful discussions.
\end{acknowledgments}

\nocite{*}

\bibliography{apssamp}

\begin{thebibliography}{75}%
\makeatletter
\providecommand \@ifxundefined [1]{%
 \@ifx{#1\undefined}
}%
\providecommand \@ifnum [1]{%
 \ifnum #1\expandafter \@firstoftwo
 \else \expandafter \@secondoftwo
 \fi
}%
\providecommand \@ifx [1]{%
 \ifx #1\expandafter \@firstoftwo
 \else \expandafter \@secondoftwo
 \fi
}%
\providecommand \natexlab [1]{#1}%
\providecommand \enquote  [1]{``#1''}%
\providecommand \bibnamefont  [1]{#1}%
\providecommand \bibfnamefont [1]{#1}%
\providecommand \citenamefont [1]{#1}%
\providecommand \href@noop [0]{\@secondoftwo}%
\providecommand \href [0]{\begingroup \@sanitize@url \@href}%
\providecommand \@href[1]{\@@startlink{#1}\@@href}%
\providecommand \@@href[1]{\endgroup#1\@@endlink}%
\providecommand \@sanitize@url [0]{\catcode `\\12\catcode `\$12\catcode
  `\&12\catcode `\#12\catcode `\^12\catcode `\_12\catcode `\%12\relax}%
\providecommand \@@startlink[1]{}%
\providecommand \@@endlink[0]{}%
\providecommand \url  [0]{\begingroup\@sanitize@url \@url }%
\providecommand \@url [1]{\endgroup\@href {#1}{\urlprefix }}%
\providecommand \urlprefix  [0]{URL }%
\providecommand \Eprint [0]{\href }%
\providecommand \doibase [0]{https://doi.org/}%
\providecommand \selectlanguage [0]{\@gobble}%
\providecommand \bibinfo  [0]{\@secondoftwo}%
\providecommand \bibfield  [0]{\@secondoftwo}%
\providecommand \translation [1]{[#1]}%
\providecommand \BibitemOpen [0]{}%
\providecommand \bibitemStop [0]{}%
\providecommand \bibitemNoStop [0]{.\EOS\space}%
\providecommand \EOS [0]{\spacefactor3000\relax}%
\providecommand \BibitemShut  [1]{\csname bibitem#1\endcsname}%
\let\auto@bib@innerbib\@empty
\bibitem [{\citenamefont {Blais}\ \emph {et~al.}(2007)\citenamefont {Blais},
  \citenamefont {Gambetta}, \citenamefont {Wallraff}, \citenamefont {Schuster},
  \citenamefont {Girvin}, \citenamefont {Devoret},\ and\ \citenamefont
  {Schoelkopf}}]{Blais2007}%
  \BibitemOpen
  \bibfield  {author} {\bibinfo {author} {\bibfnamefont {A.}~\bibnamefont
  {Blais}}, \bibinfo {author} {\bibfnamefont {J.}~\bibnamefont {Gambetta}},
  \bibinfo {author} {\bibfnamefont {A.}~\bibnamefont {Wallraff}}, \bibinfo
  {author} {\bibfnamefont {D.~I.}\ \bibnamefont {Schuster}}, \bibinfo {author}
  {\bibfnamefont {S.~M.}\ \bibnamefont {Girvin}}, \bibinfo {author}
  {\bibfnamefont {M.~H.}\ \bibnamefont {Devoret}},\ and\ \bibinfo {author}
  {\bibfnamefont {R.~J.}\ \bibnamefont {Schoelkopf}},\ }\bibfield  {title}
  {\bibinfo {title} {Quantum-information processing with circuit quantum
  electrodynamics},\ }\href {https://doi.org/10.1103/PhysRevA.75.032329}
  {\bibfield  {journal} {\bibinfo  {journal} {Phys. Rev. A}\ }\textbf {\bibinfo
  {volume} {75}},\ \bibinfo {pages} {032329} (\bibinfo {year}
  {2007})}\BibitemShut {NoStop}%
\bibitem [{\citenamefont {Schoelkopf}\ and\ \citenamefont
  {Girvin}(2008)}]{Schoelkopf2008}%
  \BibitemOpen
  \bibfield  {author} {\bibinfo {author} {\bibfnamefont {R.~J.}\ \bibnamefont
  {Schoelkopf}}\ and\ \bibinfo {author} {\bibfnamefont {S.~M.}\ \bibnamefont
  {Girvin}},\ }\bibfield  {title} {\bibinfo {title} {Wiring up quantum
  systems},\ }\href {https://doi.org/10.1038/451664a} {\bibfield  {journal}
  {\bibinfo  {journal} {Nature}\ }\textbf {\bibinfo {volume} {451}},\ \bibinfo
  {pages} {664} (\bibinfo {year} {2008})}\BibitemShut {NoStop}%
\bibitem [{\citenamefont {Reagor}\ \emph {et~al.}(2016)\citenamefont {Reagor},
  \citenamefont {Pfaff}, \citenamefont {Axline}, \citenamefont {Heeres},
  \citenamefont {Ofek}, \citenamefont {Sliwa}, \citenamefont {Holland},
  \citenamefont {Wang}, \citenamefont {Blumoff}, \citenamefont {Chou},
  \citenamefont {Hatridge}, \citenamefont {Frunzio}, \citenamefont {Devoret},
  \citenamefont {Jiang},\ and\ \citenamefont {Schoelkopf}}]{Reagor2016}%
  \BibitemOpen
  \bibfield  {author} {\bibinfo {author} {\bibfnamefont {M.}~\bibnamefont
  {Reagor}}, \bibinfo {author} {\bibfnamefont {W.}~\bibnamefont {Pfaff}},
  \bibinfo {author} {\bibfnamefont {C.}~\bibnamefont {Axline}}, \bibinfo
  {author} {\bibfnamefont {R.~W.}\ \bibnamefont {Heeres}}, \bibinfo {author}
  {\bibfnamefont {N.}~\bibnamefont {Ofek}}, \bibinfo {author} {\bibfnamefont
  {K.}~\bibnamefont {Sliwa}}, \bibinfo {author} {\bibfnamefont
  {E.}~\bibnamefont {Holland}}, \bibinfo {author} {\bibfnamefont
  {C.}~\bibnamefont {Wang}}, \bibinfo {author} {\bibfnamefont {J.}~\bibnamefont
  {Blumoff}}, \bibinfo {author} {\bibfnamefont {K.}~\bibnamefont {Chou}},
  \bibinfo {author} {\bibfnamefont {M.~J.}\ \bibnamefont {Hatridge}}, \bibinfo
  {author} {\bibfnamefont {L.}~\bibnamefont {Frunzio}}, \bibinfo {author}
  {\bibfnamefont {M.~H.}\ \bibnamefont {Devoret}}, \bibinfo {author}
  {\bibfnamefont {L.}~\bibnamefont {Jiang}},\ and\ \bibinfo {author}
  {\bibfnamefont {R.~J.}\ \bibnamefont {Schoelkopf}},\ }\bibfield  {title}
  {\bibinfo {title} {Quantum memory with millisecond coherence in circuit
  qed},\ }\href {https://doi.org/10.1103/PhysRevB.94.014506} {\bibfield
  {journal} {\bibinfo  {journal} {Phys. Rev. B}\ }\textbf {\bibinfo {volume}
  {94}},\ \bibinfo {pages} {014506} (\bibinfo {year} {2016})}\BibitemShut
  {NoStop}%
\bibitem [{\citenamefont {Hofheinz}\ \emph {et~al.}(2008)\citenamefont
  {Hofheinz}, \citenamefont {Weig}, \citenamefont {Ansmann}, \citenamefont
  {Bialczak}, \citenamefont {Lucero}, \citenamefont {Neeley}, \citenamefont
  {O'Connell}, \citenamefont {Wang}, \citenamefont {Martinis},\ and\
  \citenamefont {Cleland}}]{Hofheinz2008}%
  \BibitemOpen
  \bibfield  {author} {\bibinfo {author} {\bibfnamefont {M.}~\bibnamefont
  {Hofheinz}}, \bibinfo {author} {\bibfnamefont {E.~M.}\ \bibnamefont {Weig}},
  \bibinfo {author} {\bibfnamefont {M.}~\bibnamefont {Ansmann}}, \bibinfo
  {author} {\bibfnamefont {R.~C.}\ \bibnamefont {Bialczak}}, \bibinfo {author}
  {\bibfnamefont {E.}~\bibnamefont {Lucero}}, \bibinfo {author} {\bibfnamefont
  {M.}~\bibnamefont {Neeley}}, \bibinfo {author} {\bibfnamefont {A.~D.}\
  \bibnamefont {O'Connell}}, \bibinfo {author} {\bibfnamefont {H.}~\bibnamefont
  {Wang}}, \bibinfo {author} {\bibfnamefont {J.~M.}\ \bibnamefont {Martinis}},\
  and\ \bibinfo {author} {\bibfnamefont {A.~N.}\ \bibnamefont {Cleland}},\
  }\bibfield  {title} {\bibinfo {title} {Generation of fock states in a
  superconducting quantum circuit},\ }\href
  {https://doi.org/10.1038/nature07136} {\bibfield  {journal} {\bibinfo
  {journal} {Nature}\ }\textbf {\bibinfo {volume} {454}},\ \bibinfo {pages}
  {310} (\bibinfo {year} {2008})}\BibitemShut {NoStop}%
\bibitem [{\citenamefont {Krastanov}\ \emph {et~al.}(2015)\citenamefont
  {Krastanov}, \citenamefont {Albert}, \citenamefont {Shen}, \citenamefont
  {Zou}, \citenamefont {Heeres}, \citenamefont {Vlastakis}, \citenamefont
  {Schoelkopf},\ and\ \citenamefont {Jiang}}]{Krastanov2015}%
  \BibitemOpen
  \bibfield  {author} {\bibinfo {author} {\bibfnamefont {S.}~\bibnamefont
  {Krastanov}}, \bibinfo {author} {\bibfnamefont {V.~V.}\ \bibnamefont
  {Albert}}, \bibinfo {author} {\bibfnamefont {C.}~\bibnamefont {Shen}},
  \bibinfo {author} {\bibfnamefont {C.-L.}\ \bibnamefont {Zou}}, \bibinfo
  {author} {\bibfnamefont {R.~W.}\ \bibnamefont {Heeres}}, \bibinfo {author}
  {\bibfnamefont {B.}~\bibnamefont {Vlastakis}}, \bibinfo {author}
  {\bibfnamefont {R.~J.}\ \bibnamefont {Schoelkopf}},\ and\ \bibinfo {author}
  {\bibfnamefont {L.}~\bibnamefont {Jiang}},\ }\bibfield  {title} {\bibinfo
  {title} {Universal control of an oscillator with dispersive coupling to a
  qubit},\ }\href {https://doi.org/10.1103/PhysRevA.92.040303} {\bibfield
  {journal} {\bibinfo  {journal} {Phys. Rev. A}\ }\textbf {\bibinfo {volume}
  {92}},\ \bibinfo {pages} {040303} (\bibinfo {year} {2015})}\BibitemShut
  {NoStop}%
\bibitem [{\citenamefont {Heeres}\ \emph {et~al.}(2017)\citenamefont {Heeres},
  \citenamefont {Reinhold}, \citenamefont {Ofek}, \citenamefont {Frunzio},
  \citenamefont {Jiang}, \citenamefont {Devoret},\ and\ \citenamefont
  {Schoelkopf}}]{Heeres2017}%
  \BibitemOpen
  \bibfield  {author} {\bibinfo {author} {\bibfnamefont {R.~W.}\ \bibnamefont
  {Heeres}}, \bibinfo {author} {\bibfnamefont {P.}~\bibnamefont {Reinhold}},
  \bibinfo {author} {\bibfnamefont {N.}~\bibnamefont {Ofek}}, \bibinfo {author}
  {\bibfnamefont {L.}~\bibnamefont {Frunzio}}, \bibinfo {author} {\bibfnamefont
  {L.}~\bibnamefont {Jiang}}, \bibinfo {author} {\bibfnamefont {M.~H.}\
  \bibnamefont {Devoret}},\ and\ \bibinfo {author} {\bibfnamefont {R.~J.}\
  \bibnamefont {Schoelkopf}},\ }\bibfield  {title} {\bibinfo {title}
  {Implementing a universal gate set on a logical qubit encoded in an
  oscillator},\ }\href {https://doi.org/10.1038/s41467-017-00045-1} {\bibfield
  {journal} {\bibinfo  {journal} {Nature Communications}\ }\textbf {\bibinfo
  {volume} {8}},\ \bibinfo {pages} {94} (\bibinfo {year} {2017})}\BibitemShut
  {NoStop}%
\bibitem [{\citenamefont {Sun}\ \emph {et~al.}(2014)\citenamefont {Sun},
  \citenamefont {Petrenko}, \citenamefont {Leghtas}, \citenamefont {Vlastakis},
  \citenamefont {Kirchmair}, \citenamefont {Sliwa}, \citenamefont {Narla},
  \citenamefont {Hatridge}, \citenamefont {Shankar}, \citenamefont {Blumoff},
  \citenamefont {Frunzio}, \citenamefont {Mirrahimi}, \citenamefont {Devoret},\
  and\ \citenamefont {Schoelkopf}}]{Sun2014}%
  \BibitemOpen
  \bibfield  {author} {\bibinfo {author} {\bibfnamefont {L.}~\bibnamefont
  {Sun}}, \bibinfo {author} {\bibfnamefont {A.}~\bibnamefont {Petrenko}},
  \bibinfo {author} {\bibfnamefont {Z.}~\bibnamefont {Leghtas}}, \bibinfo
  {author} {\bibfnamefont {B.}~\bibnamefont {Vlastakis}}, \bibinfo {author}
  {\bibfnamefont {G.}~\bibnamefont {Kirchmair}}, \bibinfo {author}
  {\bibfnamefont {K.~M.}\ \bibnamefont {Sliwa}}, \bibinfo {author}
  {\bibfnamefont {A.}~\bibnamefont {Narla}}, \bibinfo {author} {\bibfnamefont
  {M.}~\bibnamefont {Hatridge}}, \bibinfo {author} {\bibfnamefont
  {S.}~\bibnamefont {Shankar}}, \bibinfo {author} {\bibfnamefont
  {J.}~\bibnamefont {Blumoff}}, \bibinfo {author} {\bibfnamefont
  {L.}~\bibnamefont {Frunzio}}, \bibinfo {author} {\bibfnamefont
  {M.}~\bibnamefont {Mirrahimi}}, \bibinfo {author} {\bibfnamefont {M.~H.}\
  \bibnamefont {Devoret}},\ and\ \bibinfo {author} {\bibfnamefont {R.~J.}\
  \bibnamefont {Schoelkopf}},\ }\bibfield  {title} {\bibinfo {title} {Tracking
  photon jumps with repeated quantum non-demolition parity measurements},\
  }\href {https://doi.org/10.1038/nature13436} {\bibfield  {journal} {\bibinfo
  {journal} {Nature}\ }\textbf {\bibinfo {volume} {511}},\ \bibinfo {pages}
  {444} (\bibinfo {year} {2014})}\BibitemShut {NoStop}%
\bibitem [{\citenamefont {Ofek}\ \emph {et~al.}(2016)\citenamefont {Ofek},
  \citenamefont {Petrenko}, \citenamefont {Heeres}, \citenamefont {Reinhold},
  \citenamefont {Leghtas}, \citenamefont {Vlastakis}, \citenamefont {Liu},
  \citenamefont {Frunzio}, \citenamefont {Girvin}, \citenamefont {Jiang},
  \citenamefont {Mirrahimi}, \citenamefont {Devoret},\ and\ \citenamefont
  {Schoelkopf}}]{Ofek2016}%
  \BibitemOpen
  \bibfield  {author} {\bibinfo {author} {\bibfnamefont {N.}~\bibnamefont
  {Ofek}}, \bibinfo {author} {\bibfnamefont {A.}~\bibnamefont {Petrenko}},
  \bibinfo {author} {\bibfnamefont {R.}~\bibnamefont {Heeres}}, \bibinfo
  {author} {\bibfnamefont {P.}~\bibnamefont {Reinhold}}, \bibinfo {author}
  {\bibfnamefont {Z.}~\bibnamefont {Leghtas}}, \bibinfo {author} {\bibfnamefont
  {B.}~\bibnamefont {Vlastakis}}, \bibinfo {author} {\bibfnamefont
  {Y.}~\bibnamefont {Liu}}, \bibinfo {author} {\bibfnamefont {L.}~\bibnamefont
  {Frunzio}}, \bibinfo {author} {\bibfnamefont {S.~M.}\ \bibnamefont {Girvin}},
  \bibinfo {author} {\bibfnamefont {L.}~\bibnamefont {Jiang}}, \bibinfo
  {author} {\bibfnamefont {M.}~\bibnamefont {Mirrahimi}}, \bibinfo {author}
  {\bibfnamefont {M.~H.}\ \bibnamefont {Devoret}},\ and\ \bibinfo {author}
  {\bibfnamefont {R.~J.}\ \bibnamefont {Schoelkopf}},\ }\bibfield  {title}
  {\bibinfo {title} {Extending the lifetime of a quantum bit with error
  correction in superconducting circuits},\ }\href
  {https://doi.org/10.1038/nature18949} {\bibfield  {journal} {\bibinfo
  {journal} {Nature}\ }\textbf {\bibinfo {volume} {536}},\ \bibinfo {pages}
  {441} (\bibinfo {year} {2016})}\BibitemShut {NoStop}%
\bibitem [{\citenamefont {Hu}\ \emph {et~al.}(2019)\citenamefont {Hu},
  \citenamefont {Ma}, \citenamefont {Cai}, \citenamefont {Mu}, \citenamefont
  {Xu}, \citenamefont {Wang}, \citenamefont {Wu}, \citenamefont {Wang},
  \citenamefont {Song}, \citenamefont {Zou}, \citenamefont {Girvin},
  \citenamefont {Duan},\ and\ \citenamefont {Sun}}]{Hu2019}%
  \BibitemOpen
  \bibfield  {author} {\bibinfo {author} {\bibfnamefont {L.}~\bibnamefont
  {Hu}}, \bibinfo {author} {\bibfnamefont {Y.}~\bibnamefont {Ma}}, \bibinfo
  {author} {\bibfnamefont {W.}~\bibnamefont {Cai}}, \bibinfo {author}
  {\bibfnamefont {X.}~\bibnamefont {Mu}}, \bibinfo {author} {\bibfnamefont
  {Y.}~\bibnamefont {Xu}}, \bibinfo {author} {\bibfnamefont {W.}~\bibnamefont
  {Wang}}, \bibinfo {author} {\bibfnamefont {Y.}~\bibnamefont {Wu}}, \bibinfo
  {author} {\bibfnamefont {H.}~\bibnamefont {Wang}}, \bibinfo {author}
  {\bibfnamefont {Y.~P.}\ \bibnamefont {Song}}, \bibinfo {author}
  {\bibfnamefont {C.-L.}\ \bibnamefont {Zou}}, \bibinfo {author} {\bibfnamefont
  {S.~M.}\ \bibnamefont {Girvin}}, \bibinfo {author} {\bibfnamefont {L.-M.}\
  \bibnamefont {Duan}},\ and\ \bibinfo {author} {\bibfnamefont
  {L.}~\bibnamefont {Sun}},\ }\bibfield  {title} {\bibinfo {title} {Quantum
  error correction and universal gate set operation on a binomial bosonic
  logical qubit},\ }\href {https://doi.org/10.1038/s41567-018-0414-3}
  {\bibfield  {journal} {\bibinfo  {journal} {Nature Physics}\ }\textbf
  {\bibinfo {volume} {15}},\ \bibinfo {pages} {503} (\bibinfo {year}
  {2019})}\BibitemShut {NoStop}%
\bibitem [{\citenamefont {Sun}\ \emph {et~al.}(2006)\citenamefont {Sun},
  \citenamefont {Wei}, \citenamefont {Liu},\ and\ \citenamefont
  {Nori}}]{Sun2006}%
  \BibitemOpen
  \bibfield  {author} {\bibinfo {author} {\bibfnamefont {C.~P.}\ \bibnamefont
  {Sun}}, \bibinfo {author} {\bibfnamefont {L.~F.}\ \bibnamefont {Wei}},
  \bibinfo {author} {\bibfnamefont {Y.-x.}\ \bibnamefont {Liu}},\ and\ \bibinfo
  {author} {\bibfnamefont {F.}~\bibnamefont {Nori}},\ }\bibfield  {title}
  {\bibinfo {title} {Quantum transducers: Integrating transmission lines and
  nanomechanical resonators via charge qubits},\ }\href
  {https://doi.org/10.1103/PhysRevA.73.022318} {\bibfield  {journal} {\bibinfo
  {journal} {Phys. Rev. A}\ }\textbf {\bibinfo {volume} {73}},\ \bibinfo
  {pages} {022318} (\bibinfo {year} {2006})}\BibitemShut {NoStop}%
\bibitem [{\citenamefont {Mariantoni}\ \emph {et~al.}(2008)\citenamefont
  {Mariantoni}, \citenamefont {Deppe}, \citenamefont {Marx}, \citenamefont
  {Gross}, \citenamefont {Wilhelm},\ and\ \citenamefont
  {Solano}}]{Mariantoni2008}%
  \BibitemOpen
  \bibfield  {author} {\bibinfo {author} {\bibfnamefont {M.}~\bibnamefont
  {Mariantoni}}, \bibinfo {author} {\bibfnamefont {F.}~\bibnamefont {Deppe}},
  \bibinfo {author} {\bibfnamefont {A.}~\bibnamefont {Marx}}, \bibinfo {author}
  {\bibfnamefont {R.}~\bibnamefont {Gross}}, \bibinfo {author} {\bibfnamefont
  {F.~K.}\ \bibnamefont {Wilhelm}},\ and\ \bibinfo {author} {\bibfnamefont
  {E.}~\bibnamefont {Solano}},\ }\bibfield  {title} {\bibinfo {title}
  {Two-resonator circuit quantum electrodynamics: A superconducting quantum
  switch},\ }\href {https://doi.org/10.1103/PhysRevB.78.104508} {\bibfield
  {journal} {\bibinfo  {journal} {Phys. Rev. B}\ }\textbf {\bibinfo {volume}
  {78}},\ \bibinfo {pages} {104508} (\bibinfo {year} {2008})}\BibitemShut
  {NoStop}%
\bibitem [{\citenamefont {Reuther}\ \emph {et~al.}(2010)\citenamefont
  {Reuther}, \citenamefont {Zueco}, \citenamefont {Deppe}, \citenamefont
  {Hoffmann}, \citenamefont {Menzel}, \citenamefont {Wei\ss{}l}, \citenamefont
  {Mariantoni}, \citenamefont {Kohler}, \citenamefont {Marx}, \citenamefont
  {Solano}, \citenamefont {Gross},\ and\ \citenamefont
  {H\"anggi}}]{Reuther2010}%
  \BibitemOpen
  \bibfield  {author} {\bibinfo {author} {\bibfnamefont {G.~M.}\ \bibnamefont
  {Reuther}}, \bibinfo {author} {\bibfnamefont {D.}~\bibnamefont {Zueco}},
  \bibinfo {author} {\bibfnamefont {F.}~\bibnamefont {Deppe}}, \bibinfo
  {author} {\bibfnamefont {E.}~\bibnamefont {Hoffmann}}, \bibinfo {author}
  {\bibfnamefont {E.~P.}\ \bibnamefont {Menzel}}, \bibinfo {author}
  {\bibfnamefont {T.}~\bibnamefont {Wei\ss{}l}}, \bibinfo {author}
  {\bibfnamefont {M.}~\bibnamefont {Mariantoni}}, \bibinfo {author}
  {\bibfnamefont {S.}~\bibnamefont {Kohler}}, \bibinfo {author} {\bibfnamefont
  {A.}~\bibnamefont {Marx}}, \bibinfo {author} {\bibfnamefont {E.}~\bibnamefont
  {Solano}}, \bibinfo {author} {\bibfnamefont {R.}~\bibnamefont {Gross}},\ and\
  \bibinfo {author} {\bibfnamefont {P.}~\bibnamefont {H\"anggi}},\ }\bibfield
  {title} {\bibinfo {title} {Two-resonator circuit quantum electrodynamics:
  Dissipative theory},\ }\href {https://doi.org/10.1103/PhysRevB.81.144510}
  {\bibfield  {journal} {\bibinfo  {journal} {Phys. Rev. B}\ }\textbf {\bibinfo
  {volume} {81}},\ \bibinfo {pages} {144510} (\bibinfo {year}
  {2010})}\BibitemShut {NoStop}%
\bibitem [{\citenamefont {Baust}\ \emph {et~al.}(2015)\citenamefont {Baust},
  \citenamefont {Hoffmann}, \citenamefont {Haeberlein}, \citenamefont
  {Schwarz}, \citenamefont {Eder}, \citenamefont {Goetz}, \citenamefont
  {Wulschner}, \citenamefont {Xie}, \citenamefont {Zhong}, \citenamefont
  {Quijandr\'{\i}a}, \citenamefont {Peropadre}, \citenamefont {Zueco},
  \citenamefont {Garc\'{\i}a~Ripoll}, \citenamefont {Solano}, \citenamefont
  {Fedorov}, \citenamefont {Menzel}, \citenamefont {Deppe}, \citenamefont
  {Marx},\ and\ \citenamefont {Gross}}]{Baust2015}%
  \BibitemOpen
  \bibfield  {author} {\bibinfo {author} {\bibfnamefont {A.}~\bibnamefont
  {Baust}}, \bibinfo {author} {\bibfnamefont {E.}~\bibnamefont {Hoffmann}},
  \bibinfo {author} {\bibfnamefont {M.}~\bibnamefont {Haeberlein}}, \bibinfo
  {author} {\bibfnamefont {M.~J.}\ \bibnamefont {Schwarz}}, \bibinfo {author}
  {\bibfnamefont {P.}~\bibnamefont {Eder}}, \bibinfo {author} {\bibfnamefont
  {J.}~\bibnamefont {Goetz}}, \bibinfo {author} {\bibfnamefont
  {F.}~\bibnamefont {Wulschner}}, \bibinfo {author} {\bibfnamefont
  {E.}~\bibnamefont {Xie}}, \bibinfo {author} {\bibfnamefont {L.}~\bibnamefont
  {Zhong}}, \bibinfo {author} {\bibfnamefont {F.}~\bibnamefont
  {Quijandr\'{\i}a}}, \bibinfo {author} {\bibfnamefont {B.}~\bibnamefont
  {Peropadre}}, \bibinfo {author} {\bibfnamefont {D.}~\bibnamefont {Zueco}},
  \bibinfo {author} {\bibfnamefont {J.-J.}\ \bibnamefont {Garc\'{\i}a~Ripoll}},
  \bibinfo {author} {\bibfnamefont {E.}~\bibnamefont {Solano}}, \bibinfo
  {author} {\bibfnamefont {K.}~\bibnamefont {Fedorov}}, \bibinfo {author}
  {\bibfnamefont {E.~P.}\ \bibnamefont {Menzel}}, \bibinfo {author}
  {\bibfnamefont {F.}~\bibnamefont {Deppe}}, \bibinfo {author} {\bibfnamefont
  {A.}~\bibnamefont {Marx}},\ and\ \bibinfo {author} {\bibfnamefont
  {R.}~\bibnamefont {Gross}},\ }\bibfield  {title} {\bibinfo {title} {Tunable
  and switchable coupling between two superconducting resonators},\ }\href
  {https://doi.org/10.1103/PhysRevB.91.014515} {\bibfield  {journal} {\bibinfo
  {journal} {Phys. Rev. B}\ }\textbf {\bibinfo {volume} {91}},\ \bibinfo
  {pages} {014515} (\bibinfo {year} {2015})}\BibitemShut {NoStop}%
\bibitem [{\citenamefont {Peropadre}\ \emph {et~al.}(2013)\citenamefont
  {Peropadre}, \citenamefont {Zueco}, \citenamefont {Wulschner}, \citenamefont
  {Deppe}, \citenamefont {Marx}, \citenamefont {Gross},\ and\ \citenamefont
  {Garc\'{\i}a-Ripoll}}]{Peropadre2013}%
  \BibitemOpen
  \bibfield  {author} {\bibinfo {author} {\bibfnamefont {B.}~\bibnamefont
  {Peropadre}}, \bibinfo {author} {\bibfnamefont {D.}~\bibnamefont {Zueco}},
  \bibinfo {author} {\bibfnamefont {F.}~\bibnamefont {Wulschner}}, \bibinfo
  {author} {\bibfnamefont {F.}~\bibnamefont {Deppe}}, \bibinfo {author}
  {\bibfnamefont {A.}~\bibnamefont {Marx}}, \bibinfo {author} {\bibfnamefont
  {R.}~\bibnamefont {Gross}},\ and\ \bibinfo {author} {\bibfnamefont {J.~J.}\
  \bibnamefont {Garc\'{\i}a-Ripoll}},\ }\bibfield  {title} {\bibinfo {title}
  {Tunable coupling engineering between superconducting resonators: From
  sidebands to effective gauge fields},\ }\href
  {https://doi.org/10.1103/PhysRevB.87.134504} {\bibfield  {journal} {\bibinfo
  {journal} {Phys. Rev. B}\ }\textbf {\bibinfo {volume} {87}},\ \bibinfo
  {pages} {134504} (\bibinfo {year} {2013})}\BibitemShut {NoStop}%
\bibitem [{\citenamefont {Wulschner}\ \emph {et~al.}(2016)\citenamefont
  {Wulschner}, \citenamefont {Goetz}, \citenamefont {Koessel}, \citenamefont
  {Hoffmann}, \citenamefont {Baust}, \citenamefont {Eder}, \citenamefont
  {Fischer}, \citenamefont {Haeberlein}, \citenamefont {Schwarz}, \citenamefont
  {Pernpeintner}, \citenamefont {Xie}, \citenamefont {Zhong}, \citenamefont
  {Zollitsch}, \citenamefont {Peropadre}, \citenamefont {Garcia Ripoll},
  \citenamefont {Solano}, \citenamefont {Fedorov}, \citenamefont {Menzel},
  \citenamefont {Deppe}, \citenamefont {Marx},\ and\ \citenamefont
  {Gross}}]{Wulschner2016}%
  \BibitemOpen
  \bibfield  {author} {\bibinfo {author} {\bibfnamefont {F.}~\bibnamefont
  {Wulschner}}, \bibinfo {author} {\bibfnamefont {J.}~\bibnamefont {Goetz}},
  \bibinfo {author} {\bibfnamefont {F.~R.}\ \bibnamefont {Koessel}}, \bibinfo
  {author} {\bibfnamefont {E.}~\bibnamefont {Hoffmann}}, \bibinfo {author}
  {\bibfnamefont {A.}~\bibnamefont {Baust}}, \bibinfo {author} {\bibfnamefont
  {P.}~\bibnamefont {Eder}}, \bibinfo {author} {\bibfnamefont {M.}~\bibnamefont
  {Fischer}}, \bibinfo {author} {\bibfnamefont {M.}~\bibnamefont {Haeberlein}},
  \bibinfo {author} {\bibfnamefont {M.~J.}\ \bibnamefont {Schwarz}}, \bibinfo
  {author} {\bibfnamefont {M.}~\bibnamefont {Pernpeintner}}, \bibinfo {author}
  {\bibfnamefont {E.}~\bibnamefont {Xie}}, \bibinfo {author} {\bibfnamefont
  {L.}~\bibnamefont {Zhong}}, \bibinfo {author} {\bibfnamefont {C.~W.}\
  \bibnamefont {Zollitsch}}, \bibinfo {author} {\bibfnamefont {B.}~\bibnamefont
  {Peropadre}}, \bibinfo {author} {\bibfnamefont {J.-J.}\ \bibnamefont
  {Garcia Ripoll}}, \bibinfo {author} {\bibfnamefont {E.}~\bibnamefont
  {Solano}}, \bibinfo {author} {\bibfnamefont {K.~G.}\ \bibnamefont {Fedorov}},
  \bibinfo {author} {\bibfnamefont {E.~P.}\ \bibnamefont {Menzel}}, \bibinfo
  {author} {\bibfnamefont {F.}~\bibnamefont {Deppe}}, \bibinfo {author}
  {\bibfnamefont {A.}~\bibnamefont {Marx}},\ and\ \bibinfo {author}
  {\bibfnamefont {R.}~\bibnamefont {Gross}},\ }\bibfield  {title} {\bibinfo
  {title} {Tunable coupling of transmission-line microwave resonators mediated
  by an rf squid},\ }\href {https://doi.org/10.1140/epjqt/s40507-016-0048-2}
  {\bibfield  {journal} {\bibinfo  {journal} {EPJ Quantum Technology}\ }\textbf
  {\bibinfo {volume} {3}},\ \bibinfo {pages} {10} (\bibinfo {year}
  {2016})}\BibitemShut {NoStop}%
\bibitem [{\citenamefont {Raussendorf}\ and\ \citenamefont
  {Harrington}(2007)}]{Raussendorf2007}%
  \BibitemOpen
  \bibfield  {author} {\bibinfo {author} {\bibfnamefont {R.}~\bibnamefont
  {Raussendorf}}\ and\ \bibinfo {author} {\bibfnamefont {J.}~\bibnamefont
  {Harrington}},\ }\bibfield  {title} {\bibinfo {title} {Fault-tolerant quantum
  computation with high threshold in two dimensions},\ }\href
  {https://doi.org/10.1103/PhysRevLett.98.190504} {\bibfield  {journal}
  {\bibinfo  {journal} {Phys. Rev. Lett.}\ }\textbf {\bibinfo {volume} {98}},\
  \bibinfo {pages} {190504} (\bibinfo {year} {2007})}\BibitemShut {NoStop}%
\bibitem [{\citenamefont {Helmer}\ \emph {et~al.}(2009)\citenamefont {Helmer},
  \citenamefont {Mariantoni}, \citenamefont {Fowler}, \citenamefont {von
  Delft}, \citenamefont {Solano},\ and\ \citenamefont
  {Marquardt}}]{Helmer_2009}%
  \BibitemOpen
  \bibfield  {author} {\bibinfo {author} {\bibfnamefont {F.}~\bibnamefont
  {Helmer}}, \bibinfo {author} {\bibfnamefont {M.}~\bibnamefont {Mariantoni}},
  \bibinfo {author} {\bibfnamefont {A.~G.}\ \bibnamefont {Fowler}}, \bibinfo
  {author} {\bibfnamefont {J.}~\bibnamefont {von Delft}}, \bibinfo {author}
  {\bibfnamefont {E.}~\bibnamefont {Solano}},\ and\ \bibinfo {author}
  {\bibfnamefont {F.}~\bibnamefont {Marquardt}},\ }\bibfield  {title} {\bibinfo
  {title} {Cavity grid for scalable quantum computation with superconducting
  circuits},\ }\href {https://doi.org/10.1209/0295-5075/85/50007} {\bibfield
  {journal} {\bibinfo  {journal} {Europhysics Letters}\ }\textbf {\bibinfo
  {volume} {85}},\ \bibinfo {pages} {50007} (\bibinfo {year}
  {2009})}\BibitemShut {NoStop}%
\bibitem [{\citenamefont {DiVincenzo}(2009)}]{DiVincenzo_2009}%
  \BibitemOpen
  \bibfield  {author} {\bibinfo {author} {\bibfnamefont {D.~P.}\ \bibnamefont
  {DiVincenzo}},\ }\bibfield  {title} {\bibinfo {title} {Fault-tolerant
  architectures for superconducting qubits},\ }\href
  {https://doi.org/10.1088/0031-8949/2009/T137/014020} {\bibfield  {journal}
  {\bibinfo  {journal} {Physica Scripta}\ }\textbf {\bibinfo {volume} {2009}},\
  \bibinfo {pages} {014020} (\bibinfo {year} {2009})}\BibitemShut {NoStop}%
\bibitem [{\citenamefont {Johnson}\ \emph {et~al.}(2010)\citenamefont
  {Johnson}, \citenamefont {Reed}, \citenamefont {Houck}, \citenamefont
  {Schuster}, \citenamefont {Bishop}, \citenamefont {Ginossar}, \citenamefont
  {Gambetta}, \citenamefont {DiCarlo}, \citenamefont {Frunzio}, \citenamefont
  {Girvin},\ and\ \citenamefont {Schoelkopf}}]{Johnson2010}%
  \BibitemOpen
  \bibfield  {author} {\bibinfo {author} {\bibfnamefont {B.~R.}\ \bibnamefont
  {Johnson}}, \bibinfo {author} {\bibfnamefont {M.~D.}\ \bibnamefont {Reed}},
  \bibinfo {author} {\bibfnamefont {A.~A.}\ \bibnamefont {Houck}}, \bibinfo
  {author} {\bibfnamefont {D.~I.}\ \bibnamefont {Schuster}}, \bibinfo {author}
  {\bibfnamefont {L.~S.}\ \bibnamefont {Bishop}}, \bibinfo {author}
  {\bibfnamefont {E.}~\bibnamefont {Ginossar}}, \bibinfo {author}
  {\bibfnamefont {J.~M.}\ \bibnamefont {Gambetta}}, \bibinfo {author}
  {\bibfnamefont {L.}~\bibnamefont {DiCarlo}}, \bibinfo {author} {\bibfnamefont
  {L.}~\bibnamefont {Frunzio}}, \bibinfo {author} {\bibfnamefont {S.~M.}\
  \bibnamefont {Girvin}},\ and\ \bibinfo {author} {\bibfnamefont {R.~J.}\
  \bibnamefont {Schoelkopf}},\ }\bibfield  {title} {\bibinfo {title} {Quantum
  non-demolition detection of single microwave photons in a circuit},\ }\href
  {https://doi.org/10.1038/nphys1710} {\bibfield  {journal} {\bibinfo
  {journal} {Nature Physics}\ }\textbf {\bibinfo {volume} {6}},\ \bibinfo
  {pages} {663} (\bibinfo {year} {2010})}\BibitemShut {NoStop}%
\bibitem [{\citenamefont {Koch}\ \emph {et~al.}(2010)\citenamefont {Koch},
  \citenamefont {Houck}, \citenamefont {Hur},\ and\ \citenamefont
  {Girvin}}]{Koch2010}%
  \BibitemOpen
  \bibfield  {author} {\bibinfo {author} {\bibfnamefont {J.}~\bibnamefont
  {Koch}}, \bibinfo {author} {\bibfnamefont {A.~A.}\ \bibnamefont {Houck}},
  \bibinfo {author} {\bibfnamefont {K.~L.}\ \bibnamefont {Hur}},\ and\ \bibinfo
  {author} {\bibfnamefont {S.~M.}\ \bibnamefont {Girvin}},\ }\bibfield  {title}
  {\bibinfo {title} {Time-reversal-symmetry breaking in circuit-qed-based
  photon lattices},\ }\href {https://doi.org/10.1103/PhysRevA.82.043811}
  {\bibfield  {journal} {\bibinfo  {journal} {Phys. Rev. A}\ }\textbf {\bibinfo
  {volume} {82}},\ \bibinfo {pages} {043811} (\bibinfo {year}
  {2010})}\BibitemShut {NoStop}%
\bibitem [{\citenamefont {Steffen}\ \emph {et~al.}(2011)\citenamefont
  {Steffen}, \citenamefont {DiVincenzo}, \citenamefont {Chow}, \citenamefont
  {Theis},\ and\ \citenamefont {Ketchen}}]{Steffen2011}%
  \BibitemOpen
  \bibfield  {author} {\bibinfo {author} {\bibfnamefont {M.}~\bibnamefont
  {Steffen}}, \bibinfo {author} {\bibfnamefont {D.~P.}\ \bibnamefont
  {DiVincenzo}}, \bibinfo {author} {\bibfnamefont {J.~M.}\ \bibnamefont
  {Chow}}, \bibinfo {author} {\bibfnamefont {T.~N.}\ \bibnamefont {Theis}},\
  and\ \bibinfo {author} {\bibfnamefont {M.~B.}\ \bibnamefont {Ketchen}},\
  }\bibfield  {title} {\bibinfo {title} {Quantum computing: An ibm
  perspective},\ }\href {https://doi.org/10.1147/JRD.2011.2165678} {\bibfield
  {journal} {\bibinfo  {journal} {IBM Journal of Research and Development}\
  }\textbf {\bibinfo {volume} {55}},\ \bibinfo {pages} {13:1} (\bibinfo {year}
  {2011})}\BibitemShut {NoStop}%
\bibitem [{\citenamefont {Wang}\ \emph {et~al.}(2011)\citenamefont {Wang},
  \citenamefont {Mariantoni}, \citenamefont {Bialczak}, \citenamefont
  {Lenander}, \citenamefont {Lucero}, \citenamefont {Neeley}, \citenamefont
  {O'Connell}, \citenamefont {Sank}, \citenamefont {Weides}, \citenamefont
  {Wenner}, \citenamefont {Yamamoto}, \citenamefont {Yin}, \citenamefont
  {Zhao}, \citenamefont {Martinis},\ and\ \citenamefont {Cleland}}]{Wang2011}%
  \BibitemOpen
  \bibfield  {author} {\bibinfo {author} {\bibfnamefont {H.}~\bibnamefont
  {Wang}}, \bibinfo {author} {\bibfnamefont {M.}~\bibnamefont {Mariantoni}},
  \bibinfo {author} {\bibfnamefont {R.~C.}\ \bibnamefont {Bialczak}}, \bibinfo
  {author} {\bibfnamefont {M.}~\bibnamefont {Lenander}}, \bibinfo {author}
  {\bibfnamefont {E.}~\bibnamefont {Lucero}}, \bibinfo {author} {\bibfnamefont
  {M.}~\bibnamefont {Neeley}}, \bibinfo {author} {\bibfnamefont {A.~D.}\
  \bibnamefont {O'Connell}}, \bibinfo {author} {\bibfnamefont {D.}~\bibnamefont
  {Sank}}, \bibinfo {author} {\bibfnamefont {M.}~\bibnamefont {Weides}},
  \bibinfo {author} {\bibfnamefont {J.}~\bibnamefont {Wenner}}, \bibinfo
  {author} {\bibfnamefont {T.}~\bibnamefont {Yamamoto}}, \bibinfo {author}
  {\bibfnamefont {Y.}~\bibnamefont {Yin}}, \bibinfo {author} {\bibfnamefont
  {J.}~\bibnamefont {Zhao}}, \bibinfo {author} {\bibfnamefont {J.~M.}\
  \bibnamefont {Martinis}},\ and\ \bibinfo {author} {\bibfnamefont {A.~N.}\
  \bibnamefont {Cleland}},\ }\bibfield  {title} {\bibinfo {title}
  {Deterministic entanglement of photons in two superconducting microwave
  resonators},\ }\href {https://doi.org/10.1103/PhysRevLett.106.060401}
  {\bibfield  {journal} {\bibinfo  {journal} {Phys. Rev. Lett.}\ }\textbf
  {\bibinfo {volume} {106}},\ \bibinfo {pages} {060401} (\bibinfo {year}
  {2011})}\BibitemShut {NoStop}%
\bibitem [{\citenamefont {Mariantoni}\ \emph {et~al.}(2011)\citenamefont
  {Mariantoni}, \citenamefont {Wang}, \citenamefont {Bialczak}, \citenamefont
  {Lenander}, \citenamefont {Lucero}, \citenamefont {Neeley}, \citenamefont
  {O'Connell}, \citenamefont {Sank}, \citenamefont {Weides}, \citenamefont
  {Wenner}, \citenamefont {Yamamoto}, \citenamefont {Yin}, \citenamefont
  {Zhao}, \citenamefont {Martinis},\ and\ \citenamefont
  {Cleland}}]{Mariantoni2011}%
  \BibitemOpen
  \bibfield  {author} {\bibinfo {author} {\bibfnamefont {M.}~\bibnamefont
  {Mariantoni}}, \bibinfo {author} {\bibfnamefont {H.}~\bibnamefont {Wang}},
  \bibinfo {author} {\bibfnamefont {R.~C.}\ \bibnamefont {Bialczak}}, \bibinfo
  {author} {\bibfnamefont {M.}~\bibnamefont {Lenander}}, \bibinfo {author}
  {\bibfnamefont {E.}~\bibnamefont {Lucero}}, \bibinfo {author} {\bibfnamefont
  {M.}~\bibnamefont {Neeley}}, \bibinfo {author} {\bibfnamefont {A.~D.}\
  \bibnamefont {O'Connell}}, \bibinfo {author} {\bibfnamefont {D.}~\bibnamefont
  {Sank}}, \bibinfo {author} {\bibfnamefont {M.}~\bibnamefont {Weides}},
  \bibinfo {author} {\bibfnamefont {J.}~\bibnamefont {Wenner}}, \bibinfo
  {author} {\bibfnamefont {T.}~\bibnamefont {Yamamoto}}, \bibinfo {author}
  {\bibfnamefont {Y.}~\bibnamefont {Yin}}, \bibinfo {author} {\bibfnamefont
  {J.}~\bibnamefont {Zhao}}, \bibinfo {author} {\bibfnamefont {J.~M.}\
  \bibnamefont {Martinis}},\ and\ \bibinfo {author} {\bibfnamefont {A.~N.}\
  \bibnamefont {Cleland}},\ }\bibfield  {title} {\bibinfo {title} {Photon shell
  game in three-resonator circuit quantum electrodynamics},\ }\href
  {https://doi.org/10.1038/nphys1885} {\bibfield  {journal} {\bibinfo
  {journal} {Nature Physics}\ }\textbf {\bibinfo {volume} {7}},\ \bibinfo
  {pages} {287} (\bibinfo {year} {2011})}\BibitemShut {NoStop}%
\bibitem [{\citenamefont {Underwood}\ \emph {et~al.}(2012)\citenamefont
  {Underwood}, \citenamefont {Shanks}, \citenamefont {Koch},\ and\
  \citenamefont {Houck}}]{Underwood2012}%
  \BibitemOpen
  \bibfield  {author} {\bibinfo {author} {\bibfnamefont {D.~L.}\ \bibnamefont
  {Underwood}}, \bibinfo {author} {\bibfnamefont {W.~E.}\ \bibnamefont
  {Shanks}}, \bibinfo {author} {\bibfnamefont {J.}~\bibnamefont {Koch}},\ and\
  \bibinfo {author} {\bibfnamefont {A.~A.}\ \bibnamefont {Houck}},\ }\bibfield
  {title} {\bibinfo {title} {Low-disorder microwave cavity lattices for quantum
  simulation with photons},\ }\href
  {https://doi.org/10.1103/PhysRevA.86.023837} {\bibfield  {journal} {\bibinfo
  {journal} {Phys. Rev. A}\ }\textbf {\bibinfo {volume} {86}},\ \bibinfo
  {pages} {023837} (\bibinfo {year} {2012})}\BibitemShut {NoStop}%
\bibitem [{\citenamefont {Fink}\ \emph {et~al.}(2009)\citenamefont {Fink},
  \citenamefont {Bianchetti}, \citenamefont {Baur}, \citenamefont {G\"oppl},
  \citenamefont {Steffen}, \citenamefont {Filipp}, \citenamefont {Leek},
  \citenamefont {Blais},\ and\ \citenamefont {Wallraff}}]{Fink2009}%
  \BibitemOpen
  \bibfield  {author} {\bibinfo {author} {\bibfnamefont {J.~M.}\ \bibnamefont
  {Fink}}, \bibinfo {author} {\bibfnamefont {R.}~\bibnamefont {Bianchetti}},
  \bibinfo {author} {\bibfnamefont {M.}~\bibnamefont {Baur}}, \bibinfo {author}
  {\bibfnamefont {M.}~\bibnamefont {G\"oppl}}, \bibinfo {author} {\bibfnamefont
  {L.}~\bibnamefont {Steffen}}, \bibinfo {author} {\bibfnamefont
  {S.}~\bibnamefont {Filipp}}, \bibinfo {author} {\bibfnamefont {P.~J.}\
  \bibnamefont {Leek}}, \bibinfo {author} {\bibfnamefont {A.}~\bibnamefont
  {Blais}},\ and\ \bibinfo {author} {\bibfnamefont {A.}~\bibnamefont
  {Wallraff}},\ }\bibfield  {title} {\bibinfo {title} {Dressed collective qubit
  states and the tavis-cummings model in circuit qed},\ }\href
  {https://doi.org/10.1103/PhysRevLett.103.083601} {\bibfield  {journal}
  {\bibinfo  {journal} {Phys. Rev. Lett.}\ }\textbf {\bibinfo {volume} {103}},\
  \bibinfo {pages} {083601} (\bibinfo {year} {2009})}\BibitemShut {NoStop}%
\bibitem [{\citenamefont {Filipp}\ \emph {et~al.}(2011)\citenamefont {Filipp},
  \citenamefont {van Loo}, \citenamefont {Baur}, \citenamefont {Steffen},\ and\
  \citenamefont {Wallraff}}]{Filipp2011}%
  \BibitemOpen
  \bibfield  {author} {\bibinfo {author} {\bibfnamefont {S.}~\bibnamefont
  {Filipp}}, \bibinfo {author} {\bibfnamefont {A.~F.}\ \bibnamefont {van Loo}},
  \bibinfo {author} {\bibfnamefont {M.}~\bibnamefont {Baur}}, \bibinfo {author}
  {\bibfnamefont {L.}~\bibnamefont {Steffen}},\ and\ \bibinfo {author}
  {\bibfnamefont {A.}~\bibnamefont {Wallraff}},\ }\bibfield  {title} {\bibinfo
  {title} {Preparation of subradiant states using local qubit control in
  circuit qed},\ }\href {https://doi.org/10.1103/PhysRevA.84.061805} {\bibfield
   {journal} {\bibinfo  {journal} {Phys. Rev. A}\ }\textbf {\bibinfo {volume}
  {84}},\ \bibinfo {pages} {061805} (\bibinfo {year} {2011})}\BibitemShut
  {NoStop}%
\bibitem [{\citenamefont {van Loo}\ \emph {et~al.}(2013)\citenamefont {van
  Loo}, \citenamefont {Fedorov}, \citenamefont {Lalumière}, \citenamefont
  {Sanders}, \citenamefont {Blais},\ and\ \citenamefont
  {Wallraff}}]{vanLoo2013}%
  \BibitemOpen
  \bibfield  {author} {\bibinfo {author} {\bibfnamefont {A.~F.}\ \bibnamefont
  {van Loo}}, \bibinfo {author} {\bibfnamefont {A.}~\bibnamefont {Fedorov}},
  \bibinfo {author} {\bibfnamefont {K.}~\bibnamefont {Lalumière}}, \bibinfo
  {author} {\bibfnamefont {B.~C.}\ \bibnamefont {Sanders}}, \bibinfo {author}
  {\bibfnamefont {A.}~\bibnamefont {Blais}},\ and\ \bibinfo {author}
  {\bibfnamefont {A.}~\bibnamefont {Wallraff}},\ }\bibfield  {title} {\bibinfo
  {title} {Photon-mediated interactions between distant artificial atoms},\
  }\href {https://doi.org/10.1126/science.1244324} {\bibfield  {journal}
  {\bibinfo  {journal} {Science}\ }\textbf {\bibinfo {volume} {342}},\ \bibinfo
  {pages} {1494} (\bibinfo {year} {2013})},\ \Eprint
  {https://arxiv.org/abs/https://www.science.org/doi/pdf/10.1126/science.1244324}
  {https://www.science.org/doi/pdf/10.1126/science.1244324} \BibitemShut
  {NoStop}%
\bibitem [{\citenamefont {Mlynek}\ \emph {et~al.}(2014)\citenamefont {Mlynek},
  \citenamefont {Abdumalikov}, \citenamefont {Eichler},\ and\ \citenamefont
  {Wallraff}}]{Mlynek2014}%
  \BibitemOpen
  \bibfield  {author} {\bibinfo {author} {\bibfnamefont {J.~A.}\ \bibnamefont
  {Mlynek}}, \bibinfo {author} {\bibfnamefont {A.~A.}\ \bibnamefont
  {Abdumalikov}}, \bibinfo {author} {\bibfnamefont {C.}~\bibnamefont
  {Eichler}},\ and\ \bibinfo {author} {\bibfnamefont {A.}~\bibnamefont
  {Wallraff}},\ }\bibfield  {title} {\bibinfo {title} {Observation of dicke
  superradiance for two artificial atoms in a cavity with high decay rate},\
  }\href {https://doi.org/10.1038/ncomms6186} {\bibfield  {journal} {\bibinfo
  {journal} {Nature Communications}\ }\textbf {\bibinfo {volume} {5}},\
  \bibinfo {pages} {5186} (\bibinfo {year} {2014})}\BibitemShut {NoStop}%
\bibitem [{\citenamefont {Lambert}\ \emph {et~al.}(2016)\citenamefont
  {Lambert}, \citenamefont {Matsuzaki}, \citenamefont {Kakuyanagi},
  \citenamefont {Ishida}, \citenamefont {Saito},\ and\ \citenamefont
  {Nori}}]{Lambert2016}%
  \BibitemOpen
  \bibfield  {author} {\bibinfo {author} {\bibfnamefont {N.}~\bibnamefont
  {Lambert}}, \bibinfo {author} {\bibfnamefont {Y.}~\bibnamefont {Matsuzaki}},
  \bibinfo {author} {\bibfnamefont {K.}~\bibnamefont {Kakuyanagi}}, \bibinfo
  {author} {\bibfnamefont {N.}~\bibnamefont {Ishida}}, \bibinfo {author}
  {\bibfnamefont {S.}~\bibnamefont {Saito}},\ and\ \bibinfo {author}
  {\bibfnamefont {F.}~\bibnamefont {Nori}},\ }\bibfield  {title} {\bibinfo
  {title} {Superradiance with an ensemble of superconducting flux qubits},\
  }\href {https://doi.org/10.1103/PhysRevB.94.224510} {\bibfield  {journal}
  {\bibinfo  {journal} {Phys. Rev. B}\ }\textbf {\bibinfo {volume} {94}},\
  \bibinfo {pages} {224510} (\bibinfo {year} {2016})}\BibitemShut {NoStop}%
\bibitem [{\citenamefont {Blais}\ \emph {et~al.}(2003)\citenamefont {Blais},
  \citenamefont {van~den Brink},\ and\ \citenamefont {Zagoskin}}]{Blais2003}%
  \BibitemOpen
  \bibfield  {author} {\bibinfo {author} {\bibfnamefont {A.}~\bibnamefont
  {Blais}}, \bibinfo {author} {\bibfnamefont {A.~M.}\ \bibnamefont {van~den
  Brink}},\ and\ \bibinfo {author} {\bibfnamefont {A.~M.}\ \bibnamefont
  {Zagoskin}},\ }\bibfield  {title} {\bibinfo {title} {Tunable coupling of
  superconducting qubits},\ }\href
  {https://doi.org/10.1103/PhysRevLett.90.127901} {\bibfield  {journal}
  {\bibinfo  {journal} {Phys. Rev. Lett.}\ }\textbf {\bibinfo {volume} {90}},\
  \bibinfo {pages} {127901} (\bibinfo {year} {2003})}\BibitemShut {NoStop}%
\bibitem [{\citenamefont {Berkley}\ \emph {et~al.}(2003)\citenamefont
  {Berkley}, \citenamefont {Xu}, \citenamefont {Ramos}, \citenamefont {Gubrud},
  \citenamefont {Strauch}, \citenamefont {Johnson}, \citenamefont {Anderson},
  \citenamefont {Dragt}, \citenamefont {Lobb},\ and\ \citenamefont
  {Wellstood}}]{Berkley2003}%
  \BibitemOpen
  \bibfield  {author} {\bibinfo {author} {\bibfnamefont {A.~J.}\ \bibnamefont
  {Berkley}}, \bibinfo {author} {\bibfnamefont {H.}~\bibnamefont {Xu}},
  \bibinfo {author} {\bibfnamefont {R.~C.}\ \bibnamefont {Ramos}}, \bibinfo
  {author} {\bibfnamefont {M.~A.}\ \bibnamefont {Gubrud}}, \bibinfo {author}
  {\bibfnamefont {F.~W.}\ \bibnamefont {Strauch}}, \bibinfo {author}
  {\bibfnamefont {P.~R.}\ \bibnamefont {Johnson}}, \bibinfo {author}
  {\bibfnamefont {J.~R.}\ \bibnamefont {Anderson}}, \bibinfo {author}
  {\bibfnamefont {A.~J.}\ \bibnamefont {Dragt}}, \bibinfo {author}
  {\bibfnamefont {C.~J.}\ \bibnamefont {Lobb}},\ and\ \bibinfo {author}
  {\bibfnamefont {F.~C.}\ \bibnamefont {Wellstood}},\ }\bibfield  {title}
  {\bibinfo {title} {Entangled macroscopic quantum states in two
  superconducting qubits},\ }\href {https://doi.org/10.1126/science.1084528}
  {\bibfield  {journal} {\bibinfo  {journal} {Science}\ }\textbf {\bibinfo
  {volume} {300}},\ \bibinfo {pages} {1548} (\bibinfo {year} {2003})},\ \Eprint
  {https://arxiv.org/abs/https://www.science.org/doi/pdf/10.1126/science.1084528}
  {https://www.science.org/doi/pdf/10.1126/science.1084528} \BibitemShut
  {NoStop}%
\bibitem [{\citenamefont {McDermott}\ \emph {et~al.}(2005)\citenamefont
  {McDermott}, \citenamefont {Simmonds}, \citenamefont {Steffen}, \citenamefont
  {Cooper}, \citenamefont {Cicak}, \citenamefont {Osborn}, \citenamefont {Oh},
  \citenamefont {Pappas},\ and\ \citenamefont {Martinis}}]{McDermott2005}%
  \BibitemOpen
  \bibfield  {author} {\bibinfo {author} {\bibfnamefont {R.}~\bibnamefont
  {McDermott}}, \bibinfo {author} {\bibfnamefont {R.~W.}\ \bibnamefont
  {Simmonds}}, \bibinfo {author} {\bibfnamefont {M.}~\bibnamefont {Steffen}},
  \bibinfo {author} {\bibfnamefont {K.~B.}\ \bibnamefont {Cooper}}, \bibinfo
  {author} {\bibfnamefont {K.}~\bibnamefont {Cicak}}, \bibinfo {author}
  {\bibfnamefont {K.~D.}\ \bibnamefont {Osborn}}, \bibinfo {author}
  {\bibfnamefont {S.}~\bibnamefont {Oh}}, \bibinfo {author} {\bibfnamefont
  {D.~P.}\ \bibnamefont {Pappas}},\ and\ \bibinfo {author} {\bibfnamefont
  {J.~M.}\ \bibnamefont {Martinis}},\ }\bibfield  {title} {\bibinfo {title}
  {Simultaneous state measurement of coupled josephson phase qubits},\ }\href
  {https://doi.org/10.1126/science.1107572} {\bibfield  {journal} {\bibinfo
  {journal} {Science}\ }\textbf {\bibinfo {volume} {307}},\ \bibinfo {pages}
  {1299} (\bibinfo {year} {2005})},\ \Eprint
  {https://arxiv.org/abs/https://www.science.org/doi/pdf/10.1126/science.1107572}
  {https://www.science.org/doi/pdf/10.1126/science.1107572} \BibitemShut
  {NoStop}%
\bibitem [{\citenamefont {Liu}\ \emph {et~al.}(2006)\citenamefont {Liu},
  \citenamefont {Wei}, \citenamefont {Tsai},\ and\ \citenamefont
  {Nori}}]{Liu2006}%
  \BibitemOpen
  \bibfield  {author} {\bibinfo {author} {\bibfnamefont {Y.-x.}\ \bibnamefont
  {Liu}}, \bibinfo {author} {\bibfnamefont {L.~F.}\ \bibnamefont {Wei}},
  \bibinfo {author} {\bibfnamefont {J.~S.}\ \bibnamefont {Tsai}},\ and\
  \bibinfo {author} {\bibfnamefont {F.}~\bibnamefont {Nori}},\ }\bibfield
  {title} {\bibinfo {title} {Controllable coupling between flux qubits},\
  }\href {https://doi.org/10.1103/PhysRevLett.96.067003} {\bibfield  {journal}
  {\bibinfo  {journal} {Phys. Rev. Lett.}\ }\textbf {\bibinfo {volume} {96}},\
  \bibinfo {pages} {067003} (\bibinfo {year} {2006})}\BibitemShut {NoStop}%
\bibitem [{\citenamefont {Hime}\ \emph {et~al.}(2006)\citenamefont {Hime},
  \citenamefont {Reichardt}, \citenamefont {Plourde}, \citenamefont
  {Robertson}, \citenamefont {Wu}, \citenamefont {Ustinov},\ and\ \citenamefont
  {Clarke}}]{Hime2006}%
  \BibitemOpen
  \bibfield  {author} {\bibinfo {author} {\bibfnamefont {T.}~\bibnamefont
  {Hime}}, \bibinfo {author} {\bibfnamefont {P.~A.}\ \bibnamefont {Reichardt}},
  \bibinfo {author} {\bibfnamefont {B.~L.~T.}\ \bibnamefont {Plourde}},
  \bibinfo {author} {\bibfnamefont {T.~L.}\ \bibnamefont {Robertson}}, \bibinfo
  {author} {\bibfnamefont {C.-E.}\ \bibnamefont {Wu}}, \bibinfo {author}
  {\bibfnamefont {A.~V.}\ \bibnamefont {Ustinov}},\ and\ \bibinfo {author}
  {\bibfnamefont {J.}~\bibnamefont {Clarke}},\ }\bibfield  {title} {\bibinfo
  {title} {Solid-state qubits with current-controlled coupling},\ }\href
  {https://doi.org/10.1126/science.1134388} {\bibfield  {journal} {\bibinfo
  {journal} {Science}\ }\textbf {\bibinfo {volume} {314}},\ \bibinfo {pages}
  {1427} (\bibinfo {year} {2006})},\ \Eprint
  {https://arxiv.org/abs/https://www.science.org/doi/pdf/10.1126/science.1134388}
  {https://www.science.org/doi/pdf/10.1126/science.1134388} \BibitemShut
  {NoStop}%
\bibitem [{\citenamefont {van~der Ploeg}\ \emph {et~al.}(2007)\citenamefont
  {van~der Ploeg}, \citenamefont {Izmalkov}, \citenamefont {van~den Brink},
  \citenamefont {H\"ubner}, \citenamefont {Grajcar}, \citenamefont {Il'ichev},
  \citenamefont {Meyer},\ and\ \citenamefont {Zagoskin}}]{vanderPloeg2007}%
  \BibitemOpen
  \bibfield  {author} {\bibinfo {author} {\bibfnamefont {S.~H.~W.}\
  \bibnamefont {van~der Ploeg}}, \bibinfo {author} {\bibfnamefont
  {A.}~\bibnamefont {Izmalkov}}, \bibinfo {author} {\bibfnamefont {A.~M.}\
  \bibnamefont {van~den Brink}}, \bibinfo {author} {\bibfnamefont
  {U.}~\bibnamefont {H\"ubner}}, \bibinfo {author} {\bibfnamefont
  {M.}~\bibnamefont {Grajcar}}, \bibinfo {author} {\bibfnamefont
  {E.}~\bibnamefont {Il'ichev}}, \bibinfo {author} {\bibfnamefont {H.-G.}\
  \bibnamefont {Meyer}},\ and\ \bibinfo {author} {\bibfnamefont {A.~M.}\
  \bibnamefont {Zagoskin}},\ }\bibfield  {title} {\bibinfo {title}
  {Controllable coupling of superconducting flux qubits},\ }\href
  {https://doi.org/10.1103/PhysRevLett.98.057004} {\bibfield  {journal}
  {\bibinfo  {journal} {Phys. Rev. Lett.}\ }\textbf {\bibinfo {volume} {98}},\
  \bibinfo {pages} {057004} (\bibinfo {year} {2007})}\BibitemShut {NoStop}%
\bibitem [{\citenamefont {Majer}\ \emph {et~al.}(2007)\citenamefont {Majer},
  \citenamefont {Chow}, \citenamefont {Gambetta}, \citenamefont {Koch},
  \citenamefont {Johnson}, \citenamefont {Schreier}, \citenamefont {Frunzio},
  \citenamefont {Schuster}, \citenamefont {Houck}, \citenamefont {Wallraff},
  \citenamefont {Blais}, \citenamefont {Devoret}, \citenamefont {Girvin},\ and\
  \citenamefont {Schoelkopf}}]{Majer2007}%
  \BibitemOpen
  \bibfield  {author} {\bibinfo {author} {\bibfnamefont {J.}~\bibnamefont
  {Majer}}, \bibinfo {author} {\bibfnamefont {J.~M.}\ \bibnamefont {Chow}},
  \bibinfo {author} {\bibfnamefont {J.~M.}\ \bibnamefont {Gambetta}}, \bibinfo
  {author} {\bibfnamefont {J.}~\bibnamefont {Koch}}, \bibinfo {author}
  {\bibfnamefont {B.~R.}\ \bibnamefont {Johnson}}, \bibinfo {author}
  {\bibfnamefont {J.~A.}\ \bibnamefont {Schreier}}, \bibinfo {author}
  {\bibfnamefont {L.}~\bibnamefont {Frunzio}}, \bibinfo {author} {\bibfnamefont
  {D.~I.}\ \bibnamefont {Schuster}}, \bibinfo {author} {\bibfnamefont {A.~A.}\
  \bibnamefont {Houck}}, \bibinfo {author} {\bibfnamefont {A.}~\bibnamefont
  {Wallraff}}, \bibinfo {author} {\bibfnamefont {A.}~\bibnamefont {Blais}},
  \bibinfo {author} {\bibfnamefont {M.~H.}\ \bibnamefont {Devoret}}, \bibinfo
  {author} {\bibfnamefont {S.~M.}\ \bibnamefont {Girvin}},\ and\ \bibinfo
  {author} {\bibfnamefont {R.~J.}\ \bibnamefont {Schoelkopf}},\ }\bibfield
  {title} {\bibinfo {title} {Coupling superconducting qubits via a cavity
  bus},\ }\href {https://doi.org/10.1038/nature06184} {\bibfield  {journal}
  {\bibinfo  {journal} {Nature}\ }\textbf {\bibinfo {volume} {449}},\ \bibinfo
  {pages} {443} (\bibinfo {year} {2007})}\BibitemShut {NoStop}%
\bibitem [{\citenamefont {Bialczak}\ \emph {et~al.}(2011)\citenamefont
  {Bialczak}, \citenamefont {Ansmann}, \citenamefont {Hofheinz}, \citenamefont
  {Lenander}, \citenamefont {Lucero}, \citenamefont {Neeley}, \citenamefont
  {O'Connell}, \citenamefont {Sank}, \citenamefont {Wang}, \citenamefont
  {Weides}, \citenamefont {Wenner}, \citenamefont {Yamamoto}, \citenamefont
  {Cleland},\ and\ \citenamefont {Martinis}}]{Bialczak2011}%
  \BibitemOpen
  \bibfield  {author} {\bibinfo {author} {\bibfnamefont {R.~C.}\ \bibnamefont
  {Bialczak}}, \bibinfo {author} {\bibfnamefont {M.}~\bibnamefont {Ansmann}},
  \bibinfo {author} {\bibfnamefont {M.}~\bibnamefont {Hofheinz}}, \bibinfo
  {author} {\bibfnamefont {M.}~\bibnamefont {Lenander}}, \bibinfo {author}
  {\bibfnamefont {E.}~\bibnamefont {Lucero}}, \bibinfo {author} {\bibfnamefont
  {M.}~\bibnamefont {Neeley}}, \bibinfo {author} {\bibfnamefont {A.~D.}\
  \bibnamefont {O'Connell}}, \bibinfo {author} {\bibfnamefont {D.}~\bibnamefont
  {Sank}}, \bibinfo {author} {\bibfnamefont {H.}~\bibnamefont {Wang}}, \bibinfo
  {author} {\bibfnamefont {M.}~\bibnamefont {Weides}}, \bibinfo {author}
  {\bibfnamefont {J.}~\bibnamefont {Wenner}}, \bibinfo {author} {\bibfnamefont
  {T.}~\bibnamefont {Yamamoto}}, \bibinfo {author} {\bibfnamefont {A.~N.}\
  \bibnamefont {Cleland}},\ and\ \bibinfo {author} {\bibfnamefont {J.~M.}\
  \bibnamefont {Martinis}},\ }\bibfield  {title} {\bibinfo {title} {Fast
  tunable coupler for superconducting qubits},\ }\href
  {https://doi.org/10.1103/PhysRevLett.106.060501} {\bibfield  {journal}
  {\bibinfo  {journal} {Phys. Rev. Lett.}\ }\textbf {\bibinfo {volume} {106}},\
  \bibinfo {pages} {060501} (\bibinfo {year} {2011})}\BibitemShut {NoStop}%
\bibitem [{\citenamefont {Cleland}\ and\ \citenamefont
  {Geller}(2004)}]{Cleland2004}%
  \BibitemOpen
  \bibfield  {author} {\bibinfo {author} {\bibfnamefont {A.~N.}\ \bibnamefont
  {Cleland}}\ and\ \bibinfo {author} {\bibfnamefont {M.~R.}\ \bibnamefont
  {Geller}},\ }\bibfield  {title} {\bibinfo {title} {Superconducting qubit
  storage and entanglement with nanomechanical resonators},\ }\href
  {https://doi.org/10.1103/PhysRevLett.93.070501} {\bibfield  {journal}
  {\bibinfo  {journal} {Phys. Rev. Lett.}\ }\textbf {\bibinfo {volume} {93}},\
  \bibinfo {pages} {070501} (\bibinfo {year} {2004})}\BibitemShut {NoStop}%
\bibitem [{\citenamefont {Sillanp{\"a}{\"a}}\ \emph {et~al.}(2007)\citenamefont
  {Sillanp{\"a}{\"a}}, \citenamefont {Park},\ and\ \citenamefont
  {Simmonds}}]{Sillanpaa2007}%
  \BibitemOpen
  \bibfield  {author} {\bibinfo {author} {\bibfnamefont {M.~A.}\ \bibnamefont
  {Sillanp{\"a}{\"a}}}, \bibinfo {author} {\bibfnamefont {J.~I.}\ \bibnamefont
  {Park}},\ and\ \bibinfo {author} {\bibfnamefont {R.~W.}\ \bibnamefont
  {Simmonds}},\ }\bibfield  {title} {\bibinfo {title} {Coherent quantum state
  storage and transfer between two phase qubits via a resonant cavity},\ }\href
  {https://doi.org/10.1038/nature06124} {\bibfield  {journal} {\bibinfo
  {journal} {Nature}\ }\textbf {\bibinfo {volume} {449}},\ \bibinfo {pages}
  {438} (\bibinfo {year} {2007})}\BibitemShut {NoStop}%
\bibitem [{\citenamefont {Leek}\ \emph {et~al.}(2010)\citenamefont {Leek},
  \citenamefont {Baur}, \citenamefont {Fink}, \citenamefont {Bianchetti},
  \citenamefont {Steffen}, \citenamefont {Filipp},\ and\ \citenamefont
  {Wallraff}}]{Leek2010}%
  \BibitemOpen
  \bibfield  {author} {\bibinfo {author} {\bibfnamefont {P.~J.}\ \bibnamefont
  {Leek}}, \bibinfo {author} {\bibfnamefont {M.}~\bibnamefont {Baur}}, \bibinfo
  {author} {\bibfnamefont {J.~M.}\ \bibnamefont {Fink}}, \bibinfo {author}
  {\bibfnamefont {R.}~\bibnamefont {Bianchetti}}, \bibinfo {author}
  {\bibfnamefont {L.}~\bibnamefont {Steffen}}, \bibinfo {author} {\bibfnamefont
  {S.}~\bibnamefont {Filipp}},\ and\ \bibinfo {author} {\bibfnamefont
  {A.}~\bibnamefont {Wallraff}},\ }\bibfield  {title} {\bibinfo {title} {Cavity
  quantum electrodynamics with separate photon storage and qubit readout
  modes},\ }\href {https://doi.org/10.1103/PhysRevLett.104.100504} {\bibfield
  {journal} {\bibinfo  {journal} {Phys. Rev. Lett.}\ }\textbf {\bibinfo
  {volume} {104}},\ \bibinfo {pages} {100504} (\bibinfo {year}
  {2010})}\BibitemShut {NoStop}%
\bibitem [{\citenamefont {Eichler}\ \emph {et~al.}(2012)\citenamefont
  {Eichler}, \citenamefont {Lang}, \citenamefont {Fink}, \citenamefont
  {Govenius}, \citenamefont {Filipp},\ and\ \citenamefont
  {Wallraff}}]{Eichler2012}%
  \BibitemOpen
  \bibfield  {author} {\bibinfo {author} {\bibfnamefont {C.}~\bibnamefont
  {Eichler}}, \bibinfo {author} {\bibfnamefont {C.}~\bibnamefont {Lang}},
  \bibinfo {author} {\bibfnamefont {J.~M.}\ \bibnamefont {Fink}}, \bibinfo
  {author} {\bibfnamefont {J.}~\bibnamefont {Govenius}}, \bibinfo {author}
  {\bibfnamefont {S.}~\bibnamefont {Filipp}},\ and\ \bibinfo {author}
  {\bibfnamefont {A.}~\bibnamefont {Wallraff}},\ }\bibfield  {title} {\bibinfo
  {title} {Observation of entanglement between itinerant microwave photons and
  a superconducting qubit},\ }\href
  {https://doi.org/10.1103/PhysRevLett.109.240501} {\bibfield  {journal}
  {\bibinfo  {journal} {Phys. Rev. Lett.}\ }\textbf {\bibinfo {volume} {109}},\
  \bibinfo {pages} {240501} (\bibinfo {year} {2012})}\BibitemShut {NoStop}%
\bibitem [{\citenamefont {Allman}\ \emph {et~al.}(2014)\citenamefont {Allman},
  \citenamefont {Whittaker}, \citenamefont {Castellanos-Beltran}, \citenamefont
  {Cicak}, \citenamefont {da~Silva}, \citenamefont {DeFeo}, \citenamefont
  {Lecocq}, \citenamefont {Sirois}, \citenamefont {Teufel}, \citenamefont
  {Aumentado},\ and\ \citenamefont {Simmonds}}]{Allman2014}%
  \BibitemOpen
  \bibfield  {author} {\bibinfo {author} {\bibfnamefont {M.~S.}\ \bibnamefont
  {Allman}}, \bibinfo {author} {\bibfnamefont {J.~D.}\ \bibnamefont
  {Whittaker}}, \bibinfo {author} {\bibfnamefont {M.}~\bibnamefont
  {Castellanos-Beltran}}, \bibinfo {author} {\bibfnamefont {K.}~\bibnamefont
  {Cicak}}, \bibinfo {author} {\bibfnamefont {F.}~\bibnamefont {da~Silva}},
  \bibinfo {author} {\bibfnamefont {M.~P.}\ \bibnamefont {DeFeo}}, \bibinfo
  {author} {\bibfnamefont {F.}~\bibnamefont {Lecocq}}, \bibinfo {author}
  {\bibfnamefont {A.}~\bibnamefont {Sirois}}, \bibinfo {author} {\bibfnamefont
  {J.~D.}\ \bibnamefont {Teufel}}, \bibinfo {author} {\bibfnamefont
  {J.}~\bibnamefont {Aumentado}},\ and\ \bibinfo {author} {\bibfnamefont
  {R.~W.}\ \bibnamefont {Simmonds}},\ }\bibfield  {title} {\bibinfo {title}
  {Tunable resonant and nonresonant interactions between a phase qubit and $lc$
  resonator},\ }\href {https://doi.org/10.1103/PhysRevLett.112.123601}
  {\bibfield  {journal} {\bibinfo  {journal} {Phys. Rev. Lett.}\ }\textbf
  {\bibinfo {volume} {112}},\ \bibinfo {pages} {123601} (\bibinfo {year}
  {2014})}\BibitemShut {NoStop}%
\bibitem [{\citenamefont {Lu}\ \emph {et~al.}(2017)\citenamefont {Lu},
  \citenamefont {Chakram}, \citenamefont {Leung}, \citenamefont {Earnest},
  \citenamefont {Naik}, \citenamefont {Huang}, \citenamefont {Groszkowski},
  \citenamefont {Kapit}, \citenamefont {Koch},\ and\ \citenamefont
  {Schuster}}]{Lu2017}%
  \BibitemOpen
  \bibfield  {author} {\bibinfo {author} {\bibfnamefont {Y.}~\bibnamefont
  {Lu}}, \bibinfo {author} {\bibfnamefont {S.}~\bibnamefont {Chakram}},
  \bibinfo {author} {\bibfnamefont {N.}~\bibnamefont {Leung}}, \bibinfo
  {author} {\bibfnamefont {N.}~\bibnamefont {Earnest}}, \bibinfo {author}
  {\bibfnamefont {R.~K.}\ \bibnamefont {Naik}}, \bibinfo {author}
  {\bibfnamefont {Z.}~\bibnamefont {Huang}}, \bibinfo {author} {\bibfnamefont
  {P.}~\bibnamefont {Groszkowski}}, \bibinfo {author} {\bibfnamefont
  {E.}~\bibnamefont {Kapit}}, \bibinfo {author} {\bibfnamefont
  {J.}~\bibnamefont {Koch}},\ and\ \bibinfo {author} {\bibfnamefont {D.~I.}\
  \bibnamefont {Schuster}},\ }\bibfield  {title} {\bibinfo {title} {Universal
  stabilization of a parametrically coupled qubit},\ }\href
  {https://doi.org/10.1103/PhysRevLett.119.150502} {\bibfield  {journal}
  {\bibinfo  {journal} {Phys. Rev. Lett.}\ }\textbf {\bibinfo {volume} {119}},\
  \bibinfo {pages} {150502} (\bibinfo {year} {2017})}\BibitemShut {NoStop}%
\bibitem [{\citenamefont {Yin}\ \emph {et~al.}(2013)\citenamefont {Yin},
  \citenamefont {Chen}, \citenamefont {Sank}, \citenamefont {O'Malley},
  \citenamefont {White}, \citenamefont {Barends}, \citenamefont {Kelly},
  \citenamefont {Lucero}, \citenamefont {Mariantoni}, \citenamefont {Megrant},
  \citenamefont {Neill}, \citenamefont {Vainsencher}, \citenamefont {Wenner},
  \citenamefont {Korotkov}, \citenamefont {Cleland},\ and\ \citenamefont
  {Martinis}}]{Yin2013}%
  \BibitemOpen
  \bibfield  {author} {\bibinfo {author} {\bibfnamefont {Y.}~\bibnamefont
  {Yin}}, \bibinfo {author} {\bibfnamefont {Y.}~\bibnamefont {Chen}}, \bibinfo
  {author} {\bibfnamefont {D.}~\bibnamefont {Sank}}, \bibinfo {author}
  {\bibfnamefont {P.~J.~J.}\ \bibnamefont {O'Malley}}, \bibinfo {author}
  {\bibfnamefont {T.~C.}\ \bibnamefont {White}}, \bibinfo {author}
  {\bibfnamefont {R.}~\bibnamefont {Barends}}, \bibinfo {author} {\bibfnamefont
  {J.}~\bibnamefont {Kelly}}, \bibinfo {author} {\bibfnamefont
  {E.}~\bibnamefont {Lucero}}, \bibinfo {author} {\bibfnamefont
  {M.}~\bibnamefont {Mariantoni}}, \bibinfo {author} {\bibfnamefont
  {A.}~\bibnamefont {Megrant}}, \bibinfo {author} {\bibfnamefont
  {C.}~\bibnamefont {Neill}}, \bibinfo {author} {\bibfnamefont
  {A.}~\bibnamefont {Vainsencher}}, \bibinfo {author} {\bibfnamefont
  {J.}~\bibnamefont {Wenner}}, \bibinfo {author} {\bibfnamefont {A.~N.}\
  \bibnamefont {Korotkov}}, \bibinfo {author} {\bibfnamefont {A.~N.}\
  \bibnamefont {Cleland}},\ and\ \bibinfo {author} {\bibfnamefont {J.~M.}\
  \bibnamefont {Martinis}},\ }\bibfield  {title} {\bibinfo {title} {Catch and
  release of microwave photon states},\ }\href
  {https://doi.org/10.1103/PhysRevLett.110.107001} {\bibfield  {journal}
  {\bibinfo  {journal} {Phys. Rev. Lett.}\ }\textbf {\bibinfo {volume} {110}},\
  \bibinfo {pages} {107001} (\bibinfo {year} {2013})}\BibitemShut {NoStop}%
\bibitem [{\citenamefont {Naik}\ \emph {et~al.}(2017)\citenamefont {Naik},
  \citenamefont {Leung}, \citenamefont {Chakram}, \citenamefont {Groszkowski},
  \citenamefont {Lu}, \citenamefont {Earnest}, \citenamefont {McKay},
  \citenamefont {Koch},\ and\ \citenamefont {Schuster}}]{Naik2017}%
  \BibitemOpen
  \bibfield  {author} {\bibinfo {author} {\bibfnamefont {R.~K.}\ \bibnamefont
  {Naik}}, \bibinfo {author} {\bibfnamefont {N.}~\bibnamefont {Leung}},
  \bibinfo {author} {\bibfnamefont {S.}~\bibnamefont {Chakram}}, \bibinfo
  {author} {\bibfnamefont {P.}~\bibnamefont {Groszkowski}}, \bibinfo {author}
  {\bibfnamefont {Y.}~\bibnamefont {Lu}}, \bibinfo {author} {\bibfnamefont
  {N.}~\bibnamefont {Earnest}}, \bibinfo {author} {\bibfnamefont {D.~C.}\
  \bibnamefont {McKay}}, \bibinfo {author} {\bibfnamefont {J.}~\bibnamefont
  {Koch}},\ and\ \bibinfo {author} {\bibfnamefont {D.~I.}\ \bibnamefont
  {Schuster}},\ }\bibfield  {title} {\bibinfo {title} {Random access quantum
  information processors using multimode circuit quantum electrodynamics},\
  }\href {https://doi.org/10.1038/s41467-017-02046-6} {\bibfield  {journal}
  {\bibinfo  {journal} {Nature Communications}\ }\textbf {\bibinfo {volume}
  {8}},\ \bibinfo {pages} {1904} (\bibinfo {year} {2017})}\BibitemShut
  {NoStop}%
\bibitem [{\citenamefont {Hann}\ \emph {et~al.}(2019)\citenamefont {Hann},
  \citenamefont {Zou}, \citenamefont {Zhang}, \citenamefont {Chu},
  \citenamefont {Schoelkopf}, \citenamefont {Girvin},\ and\ \citenamefont
  {Jiang}}]{Hann2019}%
  \BibitemOpen
  \bibfield  {author} {\bibinfo {author} {\bibfnamefont {C.~T.}\ \bibnamefont
  {Hann}}, \bibinfo {author} {\bibfnamefont {C.-L.}\ \bibnamefont {Zou}},
  \bibinfo {author} {\bibfnamefont {Y.}~\bibnamefont {Zhang}}, \bibinfo
  {author} {\bibfnamefont {Y.}~\bibnamefont {Chu}}, \bibinfo {author}
  {\bibfnamefont {R.~J.}\ \bibnamefont {Schoelkopf}}, \bibinfo {author}
  {\bibfnamefont {S.~M.}\ \bibnamefont {Girvin}},\ and\ \bibinfo {author}
  {\bibfnamefont {L.}~\bibnamefont {Jiang}},\ }\bibfield  {title} {\bibinfo
  {title} {Hardware-efficient quantum random access memory with hybrid quantum
  acoustic systems},\ }\href {https://doi.org/10.1103/PhysRevLett.123.250501}
  {\bibfield  {journal} {\bibinfo  {journal} {Phys. Rev. Lett.}\ }\textbf
  {\bibinfo {volume} {123}},\ \bibinfo {pages} {250501} (\bibinfo {year}
  {2019})}\BibitemShut {NoStop}%
\bibitem [{\citenamefont {O'Connell}\ \emph {et~al.}(2010)\citenamefont
  {O'Connell}, \citenamefont {Hofheinz}, \citenamefont {Ansmann}, \citenamefont
  {Bialczak}, \citenamefont {Lenander}, \citenamefont {Lucero}, \citenamefont
  {Neeley}, \citenamefont {Sank}, \citenamefont {Wang}, \citenamefont {Weides},
  \citenamefont {Wenner}, \citenamefont {Martinis},\ and\ \citenamefont
  {Cleland}}]{O’Connell2010}%
  \BibitemOpen
  \bibfield  {author} {\bibinfo {author} {\bibfnamefont {A.~D.}\ \bibnamefont
  {O'Connell}}, \bibinfo {author} {\bibfnamefont {M.}~\bibnamefont {Hofheinz}},
  \bibinfo {author} {\bibfnamefont {M.}~\bibnamefont {Ansmann}}, \bibinfo
  {author} {\bibfnamefont {R.~C.}\ \bibnamefont {Bialczak}}, \bibinfo {author}
  {\bibfnamefont {M.}~\bibnamefont {Lenander}}, \bibinfo {author}
  {\bibfnamefont {E.}~\bibnamefont {Lucero}}, \bibinfo {author} {\bibfnamefont
  {M.}~\bibnamefont {Neeley}}, \bibinfo {author} {\bibfnamefont
  {D.}~\bibnamefont {Sank}}, \bibinfo {author} {\bibfnamefont {H.}~\bibnamefont
  {Wang}}, \bibinfo {author} {\bibfnamefont {M.}~\bibnamefont {Weides}},
  \bibinfo {author} {\bibfnamefont {J.}~\bibnamefont {Wenner}}, \bibinfo
  {author} {\bibfnamefont {J.~M.}\ \bibnamefont {Martinis}},\ and\ \bibinfo
  {author} {\bibfnamefont {A.~N.}\ \bibnamefont {Cleland}},\ }\bibfield
  {title} {\bibinfo {title} {Quantum ground state and single-phonon control of
  a mechanical resonator},\ }\href {https://doi.org/10.1038/nature08967}
  {\bibfield  {journal} {\bibinfo  {journal} {Nature}\ }\textbf {\bibinfo
  {volume} {464}},\ \bibinfo {pages} {697} (\bibinfo {year}
  {2010})}\BibitemShut {NoStop}%
\bibitem [{\citenamefont {Pirkkalainen}\ \emph {et~al.}(2013)\citenamefont
  {Pirkkalainen}, \citenamefont {Cho}, \citenamefont {Li}, \citenamefont
  {Paraoanu}, \citenamefont {Hakonen},\ and\ \citenamefont
  {Sillanp{\"a}{\"a}}}]{Pirkkalainen2013}%
  \BibitemOpen
  \bibfield  {author} {\bibinfo {author} {\bibfnamefont {J.-M.}\ \bibnamefont
  {Pirkkalainen}}, \bibinfo {author} {\bibfnamefont {S.~U.}\ \bibnamefont
  {Cho}}, \bibinfo {author} {\bibfnamefont {J.}~\bibnamefont {Li}}, \bibinfo
  {author} {\bibfnamefont {G.~S.}\ \bibnamefont {Paraoanu}}, \bibinfo {author}
  {\bibfnamefont {P.~J.}\ \bibnamefont {Hakonen}},\ and\ \bibinfo {author}
  {\bibfnamefont {M.~A.}\ \bibnamefont {Sillanp{\"a}{\"a}}},\ }\bibfield
  {title} {\bibinfo {title} {Hybrid circuit cavity quantum electrodynamics with
  a micromechanical resonator},\ }\href {https://doi.org/10.1038/nature11821}
  {\bibfield  {journal} {\bibinfo  {journal} {Nature}\ }\textbf {\bibinfo
  {volume} {494}},\ \bibinfo {pages} {211} (\bibinfo {year}
  {2013})}\BibitemShut {NoStop}%
\bibitem [{\citenamefont {Gustafsson}\ \emph {et~al.}(2014)\citenamefont
  {Gustafsson}, \citenamefont {Aref}, \citenamefont {Kockum}, \citenamefont
  {Ekström}, \citenamefont {Johansson},\ and\ \citenamefont
  {Delsing}}]{Gustafsson2014}%
  \BibitemOpen
  \bibfield  {author} {\bibinfo {author} {\bibfnamefont {M.~V.}\ \bibnamefont
  {Gustafsson}}, \bibinfo {author} {\bibfnamefont {T.}~\bibnamefont {Aref}},
  \bibinfo {author} {\bibfnamefont {A.~F.}\ \bibnamefont {Kockum}}, \bibinfo
  {author} {\bibfnamefont {M.~K.}\ \bibnamefont {Ekström}}, \bibinfo {author}
  {\bibfnamefont {G.}~\bibnamefont {Johansson}},\ and\ \bibinfo {author}
  {\bibfnamefont {P.}~\bibnamefont {Delsing}},\ }\bibfield  {title} {\bibinfo
  {title} {Propagating phonons coupled to an artificial atom},\ }\href
  {https://doi.org/10.1126/science.1257219} {\bibfield  {journal} {\bibinfo
  {journal} {Science}\ }\textbf {\bibinfo {volume} {346}},\ \bibinfo {pages}
  {207} (\bibinfo {year} {2014})},\ \Eprint
  {https://arxiv.org/abs/https://www.science.org/doi/pdf/10.1126/science.1257219}
  {https://www.science.org/doi/pdf/10.1126/science.1257219} \BibitemShut
  {NoStop}%
\bibitem [{\citenamefont {Chu}\ \emph {et~al.}(2017)\citenamefont {Chu},
  \citenamefont {Kharel}, \citenamefont {Renninger}, \citenamefont {Burkhart},
  \citenamefont {Frunzio}, \citenamefont {Rakich},\ and\ \citenamefont
  {Schoelkopf}}]{Chu2017}%
  \BibitemOpen
  \bibfield  {author} {\bibinfo {author} {\bibfnamefont {Y.}~\bibnamefont
  {Chu}}, \bibinfo {author} {\bibfnamefont {P.}~\bibnamefont {Kharel}},
  \bibinfo {author} {\bibfnamefont {W.~H.}\ \bibnamefont {Renninger}}, \bibinfo
  {author} {\bibfnamefont {L.~D.}\ \bibnamefont {Burkhart}}, \bibinfo {author}
  {\bibfnamefont {L.}~\bibnamefont {Frunzio}}, \bibinfo {author} {\bibfnamefont
  {P.~T.}\ \bibnamefont {Rakich}},\ and\ \bibinfo {author} {\bibfnamefont
  {R.~J.}\ \bibnamefont {Schoelkopf}},\ }\bibfield  {title} {\bibinfo {title}
  {Quantum acoustics with superconducting qubits},\ }\href
  {https://doi.org/10.1126/science.aao1511} {\bibfield  {journal} {\bibinfo
  {journal} {Science}\ }\textbf {\bibinfo {volume} {358}},\ \bibinfo {pages}
  {199} (\bibinfo {year} {2017})},\ \Eprint
  {https://arxiv.org/abs/https://www.science.org/doi/pdf/10.1126/science.aao1511}
  {https://www.science.org/doi/pdf/10.1126/science.aao1511} \BibitemShut
  {NoStop}%
\bibitem [{\citenamefont {Chu}\ \emph {et~al.}(2018)\citenamefont {Chu},
  \citenamefont {Kharel}, \citenamefont {Yoon}, \citenamefont {Frunzio},
  \citenamefont {Rakich},\ and\ \citenamefont {Schoelkopf}}]{Chu2018}%
  \BibitemOpen
  \bibfield  {author} {\bibinfo {author} {\bibfnamefont {Y.}~\bibnamefont
  {Chu}}, \bibinfo {author} {\bibfnamefont {P.}~\bibnamefont {Kharel}},
  \bibinfo {author} {\bibfnamefont {T.}~\bibnamefont {Yoon}}, \bibinfo {author}
  {\bibfnamefont {L.}~\bibnamefont {Frunzio}}, \bibinfo {author} {\bibfnamefont
  {P.~T.}\ \bibnamefont {Rakich}},\ and\ \bibinfo {author} {\bibfnamefont
  {R.~J.}\ \bibnamefont {Schoelkopf}},\ }\bibfield  {title} {\bibinfo {title}
  {Creation and control of multi-phonon fock states in a bulk acoustic-wave
  resonator},\ }\href {https://doi.org/10.1038/s41586-018-0717-7} {\bibfield
  {journal} {\bibinfo  {journal} {Nature}\ }\textbf {\bibinfo {volume} {563}},\
  \bibinfo {pages} {666} (\bibinfo {year} {2018})}\BibitemShut {NoStop}%
\bibitem [{\citenamefont {Kervinen}\ \emph {et~al.}(2018)\citenamefont
  {Kervinen}, \citenamefont {Rissanen},\ and\ \citenamefont
  {Sillanp\"a\"a}}]{Kervinen2018}%
  \BibitemOpen
  \bibfield  {author} {\bibinfo {author} {\bibfnamefont {M.}~\bibnamefont
  {Kervinen}}, \bibinfo {author} {\bibfnamefont {I.}~\bibnamefont {Rissanen}},\
  and\ \bibinfo {author} {\bibfnamefont {M.}~\bibnamefont {Sillanp\"a\"a}},\
  }\bibfield  {title} {\bibinfo {title} {Interfacing planar superconducting
  qubits with high overtone bulk acoustic phonons},\ }\href
  {https://doi.org/10.1103/PhysRevB.97.205443} {\bibfield  {journal} {\bibinfo
  {journal} {Phys. Rev. B}\ }\textbf {\bibinfo {volume} {97}},\ \bibinfo
  {pages} {205443} (\bibinfo {year} {2018})}\BibitemShut {NoStop}%
\bibitem [{\citenamefont {Manenti}\ \emph {et~al.}(2017)\citenamefont
  {Manenti}, \citenamefont {Kockum}, \citenamefont {Patterson}, \citenamefont
  {Behrle}, \citenamefont {Rahamim}, \citenamefont {Tancredi}, \citenamefont
  {Nori},\ and\ \citenamefont {Leek}}]{Manenti2017}%
  \BibitemOpen
  \bibfield  {author} {\bibinfo {author} {\bibfnamefont {R.}~\bibnamefont
  {Manenti}}, \bibinfo {author} {\bibfnamefont {A.~F.}\ \bibnamefont {Kockum}},
  \bibinfo {author} {\bibfnamefont {A.}~\bibnamefont {Patterson}}, \bibinfo
  {author} {\bibfnamefont {T.}~\bibnamefont {Behrle}}, \bibinfo {author}
  {\bibfnamefont {J.}~\bibnamefont {Rahamim}}, \bibinfo {author} {\bibfnamefont
  {G.}~\bibnamefont {Tancredi}}, \bibinfo {author} {\bibfnamefont
  {F.}~\bibnamefont {Nori}},\ and\ \bibinfo {author} {\bibfnamefont {P.~J.}\
  \bibnamefont {Leek}},\ }\bibfield  {title} {\bibinfo {title} {Circuit quantum
  acoustodynamics with surface acoustic waves},\ }\href
  {https://doi.org/10.1038/s41467-017-01063-9} {\bibfield  {journal} {\bibinfo
  {journal} {Nature Communications}\ }\textbf {\bibinfo {volume} {8}},\
  \bibinfo {pages} {975} (\bibinfo {year} {2017})}\BibitemShut {NoStop}%
\bibitem [{\citenamefont {Noguchi}\ \emph {et~al.}(2017)\citenamefont
  {Noguchi}, \citenamefont {Yamazaki}, \citenamefont {Tabuchi},\ and\
  \citenamefont {Nakamura}}]{Noguchi2017}%
  \BibitemOpen
  \bibfield  {author} {\bibinfo {author} {\bibfnamefont {A.}~\bibnamefont
  {Noguchi}}, \bibinfo {author} {\bibfnamefont {R.}~\bibnamefont {Yamazaki}},
  \bibinfo {author} {\bibfnamefont {Y.}~\bibnamefont {Tabuchi}},\ and\ \bibinfo
  {author} {\bibfnamefont {Y.}~\bibnamefont {Nakamura}},\ }\bibfield  {title}
  {\bibinfo {title} {Qubit-assisted transduction for a detection of surface
  acoustic waves near the quantum limit},\ }\href
  {https://doi.org/10.1103/PhysRevLett.119.180505} {\bibfield  {journal}
  {\bibinfo  {journal} {Phys. Rev. Lett.}\ }\textbf {\bibinfo {volume} {119}},\
  \bibinfo {pages} {180505} (\bibinfo {year} {2017})}\BibitemShut {NoStop}%
\bibitem [{\citenamefont {Satzinger}\ \emph {et~al.}(2018)\citenamefont
  {Satzinger}, \citenamefont {Zhong}, \citenamefont {Chang}, \citenamefont
  {Peairs}, \citenamefont {Bienfait}, \citenamefont {Chou}, \citenamefont
  {Cleland}, \citenamefont {Conner}, \citenamefont {Dumur}, \citenamefont
  {Grebel}, \citenamefont {Gutierrez}, \citenamefont {November}, \citenamefont
  {Povey}, \citenamefont {Whiteley}, \citenamefont {Awschalom}, \citenamefont
  {Schuster},\ and\ \citenamefont {Cleland}}]{Satzinger2018}%
  \BibitemOpen
  \bibfield  {author} {\bibinfo {author} {\bibfnamefont {K.~J.}\ \bibnamefont
  {Satzinger}}, \bibinfo {author} {\bibfnamefont {Y.~P.}\ \bibnamefont
  {Zhong}}, \bibinfo {author} {\bibfnamefont {H.-S.}\ \bibnamefont {Chang}},
  \bibinfo {author} {\bibfnamefont {G.~A.}\ \bibnamefont {Peairs}}, \bibinfo
  {author} {\bibfnamefont {A.}~\bibnamefont {Bienfait}}, \bibinfo {author}
  {\bibfnamefont {M.-H.}\ \bibnamefont {Chou}}, \bibinfo {author}
  {\bibfnamefont {A.~Y.}\ \bibnamefont {Cleland}}, \bibinfo {author}
  {\bibfnamefont {C.~R.}\ \bibnamefont {Conner}}, \bibinfo {author}
  {\bibfnamefont {{\'E}.}~\bibnamefont {Dumur}}, \bibinfo {author}
  {\bibfnamefont {J.}~\bibnamefont {Grebel}}, \bibinfo {author} {\bibfnamefont
  {I.}~\bibnamefont {Gutierrez}}, \bibinfo {author} {\bibfnamefont {B.~H.}\
  \bibnamefont {November}}, \bibinfo {author} {\bibfnamefont {R.~G.}\
  \bibnamefont {Povey}}, \bibinfo {author} {\bibfnamefont {S.~J.}\ \bibnamefont
  {Whiteley}}, \bibinfo {author} {\bibfnamefont {D.~D.}\ \bibnamefont
  {Awschalom}}, \bibinfo {author} {\bibfnamefont {D.~I.}\ \bibnamefont
  {Schuster}},\ and\ \bibinfo {author} {\bibfnamefont {A.~N.}\ \bibnamefont
  {Cleland}},\ }\bibfield  {title} {\bibinfo {title} {Quantum control of
  surface acoustic-wave phonons},\ }\href
  {https://doi.org/10.1038/s41586-018-0719-5} {\bibfield  {journal} {\bibinfo
  {journal} {Nature}\ }\textbf {\bibinfo {volume} {563}},\ \bibinfo {pages}
  {661} (\bibinfo {year} {2018})}\BibitemShut {NoStop}%
\bibitem [{\citenamefont {Moores}\ \emph {et~al.}(2018)\citenamefont {Moores},
  \citenamefont {Sletten}, \citenamefont {Viennot},\ and\ \citenamefont
  {Lehnert}}]{Moores2018}%
  \BibitemOpen
  \bibfield  {author} {\bibinfo {author} {\bibfnamefont {B.~A.}\ \bibnamefont
  {Moores}}, \bibinfo {author} {\bibfnamefont {L.~R.}\ \bibnamefont {Sletten}},
  \bibinfo {author} {\bibfnamefont {J.~J.}\ \bibnamefont {Viennot}},\ and\
  \bibinfo {author} {\bibfnamefont {K.~W.}\ \bibnamefont {Lehnert}},\
  }\bibfield  {title} {\bibinfo {title} {Cavity quantum acoustic device in the
  multimode strong coupling regime},\ }\href
  {https://doi.org/10.1103/PhysRevLett.120.227701} {\bibfield  {journal}
  {\bibinfo  {journal} {Phys. Rev. Lett.}\ }\textbf {\bibinfo {volume} {120}},\
  \bibinfo {pages} {227701} (\bibinfo {year} {2018})}\BibitemShut {NoStop}%
\bibitem [{\citenamefont {Bolgar}\ \emph {et~al.}(2018)\citenamefont {Bolgar},
  \citenamefont {Zotova}, \citenamefont {Kirichenko}, \citenamefont {Besedin},
  \citenamefont {Semenov}, \citenamefont {Shaikhaidarov},\ and\ \citenamefont
  {Astafiev}}]{Bolgar2018}%
  \BibitemOpen
  \bibfield  {author} {\bibinfo {author} {\bibfnamefont {A.~N.}\ \bibnamefont
  {Bolgar}}, \bibinfo {author} {\bibfnamefont {J.~I.}\ \bibnamefont {Zotova}},
  \bibinfo {author} {\bibfnamefont {D.~D.}\ \bibnamefont {Kirichenko}},
  \bibinfo {author} {\bibfnamefont {I.~S.}\ \bibnamefont {Besedin}}, \bibinfo
  {author} {\bibfnamefont {A.~V.}\ \bibnamefont {Semenov}}, \bibinfo {author}
  {\bibfnamefont {R.~S.}\ \bibnamefont {Shaikhaidarov}},\ and\ \bibinfo
  {author} {\bibfnamefont {O.~V.}\ \bibnamefont {Astafiev}},\ }\bibfield
  {title} {\bibinfo {title} {Quantum regime of a two-dimensional phonon
  cavity},\ }\href {https://doi.org/10.1103/PhysRevLett.120.223603} {\bibfield
  {journal} {\bibinfo  {journal} {Phys. Rev. Lett.}\ }\textbf {\bibinfo
  {volume} {120}},\ \bibinfo {pages} {223603} (\bibinfo {year}
  {2018})}\BibitemShut {NoStop}%
\bibitem [{\citenamefont {Sletten}\ \emph {et~al.}(2019)\citenamefont
  {Sletten}, \citenamefont {Moores}, \citenamefont {Viennot},\ and\
  \citenamefont {Lehnert}}]{Sletten2019}%
  \BibitemOpen
  \bibfield  {author} {\bibinfo {author} {\bibfnamefont {L.~R.}\ \bibnamefont
  {Sletten}}, \bibinfo {author} {\bibfnamefont {B.~A.}\ \bibnamefont {Moores}},
  \bibinfo {author} {\bibfnamefont {J.~J.}\ \bibnamefont {Viennot}},\ and\
  \bibinfo {author} {\bibfnamefont {K.~W.}\ \bibnamefont {Lehnert}},\
  }\bibfield  {title} {\bibinfo {title} {Resolving phonon fock states in a
  multimode cavity with a double-slit qubit},\ }\href
  {https://doi.org/10.1103/PhysRevX.9.021056} {\bibfield  {journal} {\bibinfo
  {journal} {Phys. Rev. X}\ }\textbf {\bibinfo {volume} {9}},\ \bibinfo {pages}
  {021056} (\bibinfo {year} {2019})}\BibitemShut {NoStop}%
\bibitem [{\citenamefont {Arrangoiz-Arriola}\ \emph {et~al.}(2019)\citenamefont
  {Arrangoiz-Arriola}, \citenamefont {Wollack}, \citenamefont {Wang},
  \citenamefont {Pechal}, \citenamefont {Jiang}, \citenamefont {McKenna},
  \citenamefont {Witmer}, \citenamefont {Van~Laer},\ and\ \citenamefont
  {Safavi-Naeini}}]{Arrangoiz-Arriola2019}%
  \BibitemOpen
  \bibfield  {author} {\bibinfo {author} {\bibfnamefont {P.}~\bibnamefont
  {Arrangoiz-Arriola}}, \bibinfo {author} {\bibfnamefont {E.~A.}\ \bibnamefont
  {Wollack}}, \bibinfo {author} {\bibfnamefont {Z.}~\bibnamefont {Wang}},
  \bibinfo {author} {\bibfnamefont {M.}~\bibnamefont {Pechal}}, \bibinfo
  {author} {\bibfnamefont {W.}~\bibnamefont {Jiang}}, \bibinfo {author}
  {\bibfnamefont {T.~P.}\ \bibnamefont {McKenna}}, \bibinfo {author}
  {\bibfnamefont {J.~D.}\ \bibnamefont {Witmer}}, \bibinfo {author}
  {\bibfnamefont {R.}~\bibnamefont {Van~Laer}},\ and\ \bibinfo {author}
  {\bibfnamefont {A.~H.}\ \bibnamefont {Safavi-Naeini}},\ }\bibfield  {title}
  {\bibinfo {title} {Resolving the energy levels of a nanomechanical
  oscillator},\ }\href {https://doi.org/10.1038/s41586-019-1386-x} {\bibfield
  {journal} {\bibinfo  {journal} {Nature}\ }\textbf {\bibinfo {volume} {571}},\
  \bibinfo {pages} {537} (\bibinfo {year} {2019})}\BibitemShut {NoStop}%
\bibitem [{\citenamefont {et~al.}(2013)}]{Mutus2013}%
  \BibitemOpen
  \bibfield  {author} {\bibinfo {author} {\bibfnamefont {J.~M.}\ \bibnamefont
  {et~al.}},\ }\bibfield  {title} {\bibinfo {title} {Design and
  characterization of a lumped element single-ended superconducting microwave
  parametric amplifier with on-chip flux bias line},\ }\href
  {https://aip.scitation.org/doi/10.1063/1.4821136} {\bibfield  {journal}
  {\bibinfo  {journal} {Appl. Phys. Lett.}\ }\textbf {\bibinfo {volume}
  {103}},\ \bibinfo {pages} {122602} (\bibinfo {year} {2013})}\BibitemShut
  {NoStop}%
\bibitem [{\citenamefont {Leghtas}\ \emph {et~al.}(2015)\citenamefont
  {Leghtas}, \citenamefont {Touzard}, \citenamefont {Pop}, \citenamefont {Kou},
  \citenamefont {Vlastakis}, \citenamefont {Petrenko}, \citenamefont {Sliwa},
  \citenamefont {Narla}, \citenamefont {Shankar}, \citenamefont {Hatridge},
  \citenamefont {Reagor}, \citenamefont {Frunzio}, \citenamefont {Schoelkopf},
  \citenamefont {Mirrahimi},\ and\ \citenamefont {Devoret}}]{Leghtas2015}%
  \BibitemOpen
  \bibfield  {author} {\bibinfo {author} {\bibfnamefont {Z.}~\bibnamefont
  {Leghtas}}, \bibinfo {author} {\bibfnamefont {S.}~\bibnamefont {Touzard}},
  \bibinfo {author} {\bibfnamefont {I.~M.}\ \bibnamefont {Pop}}, \bibinfo
  {author} {\bibfnamefont {A.}~\bibnamefont {Kou}}, \bibinfo {author}
  {\bibfnamefont {B.}~\bibnamefont {Vlastakis}}, \bibinfo {author}
  {\bibfnamefont {A.}~\bibnamefont {Petrenko}}, \bibinfo {author}
  {\bibfnamefont {K.~M.}\ \bibnamefont {Sliwa}}, \bibinfo {author}
  {\bibfnamefont {A.}~\bibnamefont {Narla}}, \bibinfo {author} {\bibfnamefont
  {S.}~\bibnamefont {Shankar}}, \bibinfo {author} {\bibfnamefont {M.~J.}\
  \bibnamefont {Hatridge}}, \bibinfo {author} {\bibfnamefont {M.}~\bibnamefont
  {Reagor}}, \bibinfo {author} {\bibfnamefont {L.}~\bibnamefont {Frunzio}},
  \bibinfo {author} {\bibfnamefont {R.~J.}\ \bibnamefont {Schoelkopf}},
  \bibinfo {author} {\bibfnamefont {M.}~\bibnamefont {Mirrahimi}},\ and\
  \bibinfo {author} {\bibfnamefont {M.~H.}\ \bibnamefont {Devoret}},\
  }\bibfield  {title} {\bibinfo {title} {Confining the state of light to a
  quantum manifold by engineered two-photon loss},\ }\href
  {https://doi.org/10.1126/science.aaa2085} {\bibfield  {journal} {\bibinfo
  {journal} {Science}\ }\textbf {\bibinfo {volume} {347}},\ \bibinfo {pages}
  {853} (\bibinfo {year} {2015})},\ \Eprint
  {https://arxiv.org/abs/https://www.science.org/doi/pdf/10.1126/science.aaa2085}
  {https://www.science.org/doi/pdf/10.1126/science.aaa2085} \BibitemShut
  {NoStop}%
\bibitem [{\citenamefont {Gao}\ \emph {et~al.}(2018)\citenamefont {Gao},
  \citenamefont {Lester}, \citenamefont {Zhang}, \citenamefont {Wang},
  \citenamefont {Rosenblum}, \citenamefont {Frunzio}, \citenamefont {Jiang},
  \citenamefont {Girvin},\ and\ \citenamefont {Schoelkopf}}]{Gao2018}%
  \BibitemOpen
  \bibfield  {author} {\bibinfo {author} {\bibfnamefont {Y.~Y.}\ \bibnamefont
  {Gao}}, \bibinfo {author} {\bibfnamefont {B.~J.}\ \bibnamefont {Lester}},
  \bibinfo {author} {\bibfnamefont {Y.}~\bibnamefont {Zhang}}, \bibinfo
  {author} {\bibfnamefont {C.}~\bibnamefont {Wang}}, \bibinfo {author}
  {\bibfnamefont {S.}~\bibnamefont {Rosenblum}}, \bibinfo {author}
  {\bibfnamefont {L.}~\bibnamefont {Frunzio}}, \bibinfo {author} {\bibfnamefont
  {L.}~\bibnamefont {Jiang}}, \bibinfo {author} {\bibfnamefont {S.~M.}\
  \bibnamefont {Girvin}},\ and\ \bibinfo {author} {\bibfnamefont {R.~J.}\
  \bibnamefont {Schoelkopf}},\ }\bibfield  {title} {\bibinfo {title}
  {Programmable interference between two microwave quantum memories},\ }\href
  {https://doi.org/10.1103/PhysRevX.8.021073} {\bibfield  {journal} {\bibinfo
  {journal} {Phys. Rev. X}\ }\textbf {\bibinfo {volume} {8}},\ \bibinfo {pages}
  {021073} (\bibinfo {year} {2018})}\BibitemShut {NoStop}%
\bibitem [{\citenamefont {Zhang}\ \emph {et~al.}(2019)\citenamefont {Zhang},
  \citenamefont {Lester}, \citenamefont {Gao}, \citenamefont {Jiang},
  \citenamefont {Schoelkopf},\ and\ \citenamefont {Girvin}}]{Zhang2019}%
  \BibitemOpen
  \bibfield  {author} {\bibinfo {author} {\bibfnamefont {Y.}~\bibnamefont
  {Zhang}}, \bibinfo {author} {\bibfnamefont {B.~J.}\ \bibnamefont {Lester}},
  \bibinfo {author} {\bibfnamefont {Y.~Y.}\ \bibnamefont {Gao}}, \bibinfo
  {author} {\bibfnamefont {L.}~\bibnamefont {Jiang}}, \bibinfo {author}
  {\bibfnamefont {R.~J.}\ \bibnamefont {Schoelkopf}},\ and\ \bibinfo {author}
  {\bibfnamefont {S.~M.}\ \bibnamefont {Girvin}},\ }\bibfield  {title}
  {\bibinfo {title} {Engineering bilinear mode coupling in circuit qed: Theory
  and experiment},\ }\href {https://doi.org/10.1103/PhysRevA.99.012314}
  {\bibfield  {journal} {\bibinfo  {journal} {Phys. Rev. A}\ }\textbf {\bibinfo
  {volume} {99}},\ \bibinfo {pages} {012314} (\bibinfo {year}
  {2019})}\BibitemShut {NoStop}%
\bibitem [{\citenamefont {Chen}\ \emph {et~al.}(2018)\citenamefont {Chen},
  \citenamefont {Liu}, \citenamefont {Sun},\ and\ \citenamefont
  {Wu}}]{Chen2018}%
  \BibitemOpen
  \bibfield  {author} {\bibinfo {author} {\bibfnamefont {Q.-M.}\ \bibnamefont
  {Chen}}, \bibinfo {author} {\bibfnamefont {Y.-x.}\ \bibnamefont {Liu}},
  \bibinfo {author} {\bibfnamefont {L.}~\bibnamefont {Sun}},\ and\ \bibinfo
  {author} {\bibfnamefont {R.-B.}\ \bibnamefont {Wu}},\ }\bibfield  {title}
  {\bibinfo {title} {Tuning the coupling between superconducting resonators
  with collective qubits},\ }\href {https://doi.org/10.1103/PhysRevA.98.042328}
  {\bibfield  {journal} {\bibinfo  {journal} {Phys. Rev. A}\ }\textbf {\bibinfo
  {volume} {98}},\ \bibinfo {pages} {042328} (\bibinfo {year}
  {2018})}\BibitemShut {NoStop}%
\bibitem [{\citenamefont {Dicke}(1954)}]{Dicke1954}%
  \BibitemOpen
  \bibfield  {author} {\bibinfo {author} {\bibfnamefont {R.~H.}\ \bibnamefont
  {Dicke}},\ }\bibfield  {title} {\bibinfo {title} {Coherence in spontaneous
  radiation processes},\ }\href {https://doi.org/10.1103/PhysRev.93.99}
  {\bibfield  {journal} {\bibinfo  {journal} {Phys. Rev.}\ }\textbf {\bibinfo
  {volume} {93}},\ \bibinfo {pages} {99} (\bibinfo {year} {1954})}\BibitemShut
  {NoStop}%
\bibitem [{\citenamefont {Gross}\ and\ \citenamefont
  {Haroche}(1982)}]{Gross1982}%
  \BibitemOpen
  \bibfield  {author} {\bibinfo {author} {\bibfnamefont {M.}~\bibnamefont
  {Gross}}\ and\ \bibinfo {author} {\bibfnamefont {S.}~\bibnamefont
  {Haroche}},\ }\bibfield  {title} {\bibinfo {title} {Superradiance: An essay
  on the theory of collective spontaneous emission},\ }\href
  {https://doi.org/https://doi.org/10.1016/0370-1573(82)90102-8} {\bibfield
  {journal} {\bibinfo  {journal} {Physics Reports}\ }\textbf {\bibinfo {volume}
  {93}},\ \bibinfo {pages} {301} (\bibinfo {year} {1982})}\BibitemShut
  {NoStop}%
\bibitem [{\citenamefont {Freedhoff}\ and\ \citenamefont
  {Kranendonk}(1967)}]{Freedhoff1967}%
  \BibitemOpen
  \bibfield  {author} {\bibinfo {author} {\bibfnamefont {H.}~\bibnamefont
  {Freedhoff}}\ and\ \bibinfo {author} {\bibfnamefont {J.~V.}\ \bibnamefont
  {Kranendonk}},\ }\bibfield  {title} {\bibinfo {title} {Theory of coherent
  resonant absorption and emission at infrared and optical frequencies},\
  }\href {https://doi.org/10.1139/p67-142} {\bibfield  {journal} {\bibinfo
  {journal} {Canadian Journal of Physics}\ }\textbf {\bibinfo {volume} {45}},\
  \bibinfo {pages} {1833} (\bibinfo {year} {1967})}\BibitemShut {NoStop}%
\bibitem [{\citenamefont {Stroud}\ \emph {et~al.}(1972)\citenamefont {Stroud},
  \citenamefont {Eberly}, \citenamefont {Lama},\ and\ \citenamefont
  {Mandel}}]{Stroud1972}%
  \BibitemOpen
  \bibfield  {author} {\bibinfo {author} {\bibfnamefont {C.~R.}\ \bibnamefont
  {Stroud}}, \bibinfo {author} {\bibfnamefont {J.~H.}\ \bibnamefont {Eberly}},
  \bibinfo {author} {\bibfnamefont {W.~L.}\ \bibnamefont {Lama}},\ and\
  \bibinfo {author} {\bibfnamefont {L.}~\bibnamefont {Mandel}},\ }\bibfield
  {title} {\bibinfo {title} {Superradiant effects in systems of two-level
  atoms},\ }\href {https://doi.org/10.1103/PhysRevA.5.1094} {\bibfield
  {journal} {\bibinfo  {journal} {Phys. Rev. A}\ }\textbf {\bibinfo {volume}
  {5}},\ \bibinfo {pages} {1094} (\bibinfo {year} {1972})}\BibitemShut
  {NoStop}%
\bibitem [{\citenamefont {Pavolini}\ \emph {et~al.}(1985)\citenamefont
  {Pavolini}, \citenamefont {Crubellier}, \citenamefont {Pillet}, \citenamefont
  {Cabaret},\ and\ \citenamefont {Liberman}}]{Pavolini1985}%
  \BibitemOpen
  \bibfield  {author} {\bibinfo {author} {\bibfnamefont {D.}~\bibnamefont
  {Pavolini}}, \bibinfo {author} {\bibfnamefont {A.}~\bibnamefont
  {Crubellier}}, \bibinfo {author} {\bibfnamefont {P.}~\bibnamefont {Pillet}},
  \bibinfo {author} {\bibfnamefont {L.}~\bibnamefont {Cabaret}},\ and\ \bibinfo
  {author} {\bibfnamefont {S.}~\bibnamefont {Liberman}},\ }\bibfield  {title}
  {\bibinfo {title} {Experimental evidence for subradiance},\ }\href
  {https://doi.org/10.1103/PhysRevLett.54.1917} {\bibfield  {journal} {\bibinfo
   {journal} {Phys. Rev. Lett.}\ }\textbf {\bibinfo {volume} {54}},\ \bibinfo
  {pages} {1917} (\bibinfo {year} {1985})}\BibitemShut {NoStop}%
\bibitem [{\citenamefont {Pechal}\ \emph {et~al.}(2014)\citenamefont {Pechal},
  \citenamefont {Huthmacher}, \citenamefont {Eichler}, \citenamefont
  {Zeytino\ifmmode~\breve{g}\else \u{g}\fi{}lu}, \citenamefont {Abdumalikov},
  \citenamefont {Berger}, \citenamefont {Wallraff},\ and\ \citenamefont
  {Filipp}}]{Pechal2014}%
  \BibitemOpen
  \bibfield  {author} {\bibinfo {author} {\bibfnamefont {M.}~\bibnamefont
  {Pechal}}, \bibinfo {author} {\bibfnamefont {L.}~\bibnamefont {Huthmacher}},
  \bibinfo {author} {\bibfnamefont {C.}~\bibnamefont {Eichler}}, \bibinfo
  {author} {\bibfnamefont {S.}~\bibnamefont {Zeytino\ifmmode~\breve{g}\else
  \u{g}\fi{}lu}}, \bibinfo {author} {\bibfnamefont {A.~A.}\ \bibnamefont
  {Abdumalikov}}, \bibinfo {author} {\bibfnamefont {S.}~\bibnamefont {Berger}},
  \bibinfo {author} {\bibfnamefont {A.}~\bibnamefont {Wallraff}},\ and\
  \bibinfo {author} {\bibfnamefont {S.}~\bibnamefont {Filipp}},\ }\bibfield
  {title} {\bibinfo {title} {Microwave-controlled generation of shaped single
  photons in circuit quantum electrodynamics},\ }\href
  {https://doi.org/10.1103/PhysRevX.4.041010} {\bibfield  {journal} {\bibinfo
  {journal} {Phys. Rev. X}\ }\textbf {\bibinfo {volume} {4}},\ \bibinfo {pages}
  {041010} (\bibinfo {year} {2014})}\BibitemShut {NoStop}%
\bibitem [{\citenamefont {Srinivasan}\ \emph {et~al.}(2014)\citenamefont
  {Srinivasan}, \citenamefont {Sundaresan}, \citenamefont {Sadri},
  \citenamefont {Liu}, \citenamefont {Gambetta}, \citenamefont {Yu},
  \citenamefont {Girvin},\ and\ \citenamefont {Houck}}]{Srinivasan2014}%
  \BibitemOpen
  \bibfield  {author} {\bibinfo {author} {\bibfnamefont {S.~J.}\ \bibnamefont
  {Srinivasan}}, \bibinfo {author} {\bibfnamefont {N.~M.}\ \bibnamefont
  {Sundaresan}}, \bibinfo {author} {\bibfnamefont {D.}~\bibnamefont {Sadri}},
  \bibinfo {author} {\bibfnamefont {Y.}~\bibnamefont {Liu}}, \bibinfo {author}
  {\bibfnamefont {J.~M.}\ \bibnamefont {Gambetta}}, \bibinfo {author}
  {\bibfnamefont {T.}~\bibnamefont {Yu}}, \bibinfo {author} {\bibfnamefont
  {S.~M.}\ \bibnamefont {Girvin}},\ and\ \bibinfo {author} {\bibfnamefont
  {A.~A.}\ \bibnamefont {Houck}},\ }\bibfield  {title} {\bibinfo {title}
  {Time-reversal symmetrization of spontaneous emission for quantum state
  transfer},\ }\href {https://doi.org/10.1103/PhysRevA.89.033857} {\bibfield
  {journal} {\bibinfo  {journal} {Phys. Rev. A}\ }\textbf {\bibinfo {volume}
  {89}},\ \bibinfo {pages} {033857} (\bibinfo {year} {2014})}\BibitemShut
  {NoStop}%
\bibitem [{\citenamefont {Axline}\ \emph {et~al.}(2018)\citenamefont {Axline},
  \citenamefont {Burkhart}, \citenamefont {Pfaff}, \citenamefont {Zhang},
  \citenamefont {Chou}, \citenamefont {Campagne-Ibarcq}, \citenamefont
  {Reinhold}, \citenamefont {Frunzio}, \citenamefont {Girvin}, \citenamefont
  {Jiang}, \citenamefont {Devoret},\ and\ \citenamefont
  {Schoelkopf}}]{Axline2018}%
  \BibitemOpen
  \bibfield  {author} {\bibinfo {author} {\bibfnamefont {C.~J.}\ \bibnamefont
  {Axline}}, \bibinfo {author} {\bibfnamefont {L.~D.}\ \bibnamefont
  {Burkhart}}, \bibinfo {author} {\bibfnamefont {W.}~\bibnamefont {Pfaff}},
  \bibinfo {author} {\bibfnamefont {M.}~\bibnamefont {Zhang}}, \bibinfo
  {author} {\bibfnamefont {K.}~\bibnamefont {Chou}}, \bibinfo {author}
  {\bibfnamefont {P.}~\bibnamefont {Campagne-Ibarcq}}, \bibinfo {author}
  {\bibfnamefont {P.}~\bibnamefont {Reinhold}}, \bibinfo {author}
  {\bibfnamefont {L.}~\bibnamefont {Frunzio}}, \bibinfo {author} {\bibfnamefont
  {S.~M.}\ \bibnamefont {Girvin}}, \bibinfo {author} {\bibfnamefont
  {L.}~\bibnamefont {Jiang}}, \bibinfo {author} {\bibfnamefont {M.~H.}\
  \bibnamefont {Devoret}},\ and\ \bibinfo {author} {\bibfnamefont {R.~J.}\
  \bibnamefont {Schoelkopf}},\ }\bibfield  {title} {\bibinfo {title} {On-demand
  quantum state transfer and entanglement between remote microwave cavity
  memories},\ }\href {https://doi.org/10.1038/s41567-018-0115-y} {\bibfield
  {journal} {\bibinfo  {journal} {Nature Physics}\ }\textbf {\bibinfo {volume}
  {14}},\ \bibinfo {pages} {705} (\bibinfo {year} {2018})}\BibitemShut
  {NoStop}%
\bibitem [{\citenamefont {Kurpiers}\ \emph {et~al.}(2018)\citenamefont
  {Kurpiers}, \citenamefont {Magnard}, \citenamefont {Walter}, \citenamefont
  {Royer}, \citenamefont {Pechal}, \citenamefont {Heinsoo}, \citenamefont
  {Salath{\'e}}, \citenamefont {Akin}, \citenamefont {Storz}, \citenamefont
  {Besse}, \citenamefont {Gasparinetti}, \citenamefont {Blais},\ and\
  \citenamefont {Wallraff}}]{Kurpiers2018}%
  \BibitemOpen
  \bibfield  {author} {\bibinfo {author} {\bibfnamefont {P.}~\bibnamefont
  {Kurpiers}}, \bibinfo {author} {\bibfnamefont {P.}~\bibnamefont {Magnard}},
  \bibinfo {author} {\bibfnamefont {T.}~\bibnamefont {Walter}}, \bibinfo
  {author} {\bibfnamefont {B.}~\bibnamefont {Royer}}, \bibinfo {author}
  {\bibfnamefont {M.}~\bibnamefont {Pechal}}, \bibinfo {author} {\bibfnamefont
  {J.}~\bibnamefont {Heinsoo}}, \bibinfo {author} {\bibfnamefont
  {Y.}~\bibnamefont {Salath{\'e}}}, \bibinfo {author} {\bibfnamefont
  {A.}~\bibnamefont {Akin}}, \bibinfo {author} {\bibfnamefont {S.}~\bibnamefont
  {Storz}}, \bibinfo {author} {\bibfnamefont {J.-C.}\ \bibnamefont {Besse}},
  \bibinfo {author} {\bibfnamefont {S.}~\bibnamefont {Gasparinetti}}, \bibinfo
  {author} {\bibfnamefont {A.}~\bibnamefont {Blais}},\ and\ \bibinfo {author}
  {\bibfnamefont {A.}~\bibnamefont {Wallraff}},\ }\bibfield  {title} {\bibinfo
  {title} {Deterministic quantum state transfer and remote entanglement using
  microwave photons},\ }\href {https://doi.org/10.1038/s41586-018-0195-y}
  {\bibfield  {journal} {\bibinfo  {journal} {Nature}\ }\textbf {\bibinfo
  {volume} {558}},\ \bibinfo {pages} {264} (\bibinfo {year}
  {2018})}\BibitemShut {NoStop}%
\bibitem [{\citenamefont {Knill}\ \emph {et~al.}(2001)\citenamefont {Knill},
  \citenamefont {Laflamme},\ and\ \citenamefont {Milburn}}]{Knill2001}%
  \BibitemOpen
  \bibfield  {author} {\bibinfo {author} {\bibfnamefont {E.}~\bibnamefont
  {Knill}}, \bibinfo {author} {\bibfnamefont {R.}~\bibnamefont {Laflamme}},\
  and\ \bibinfo {author} {\bibfnamefont {G.~J.}\ \bibnamefont {Milburn}},\
  }\bibfield  {title} {\bibinfo {title} {A scheme for efficient quantum
  computation with linear optics},\ }\href {https://doi.org/10.1038/35051009}
  {\bibfield  {journal} {\bibinfo  {journal} {Nature}\ }\textbf {\bibinfo
  {volume} {409}},\ \bibinfo {pages} {46} (\bibinfo {year} {2001})}\BibitemShut
  {NoStop}%
\bibitem [{\citenamefont {Niu}\ \emph {et~al.}(2018)\citenamefont {Niu},
  \citenamefont {Chuang},\ and\ \citenamefont {Shapiro}}]{Niu2018}%
  \BibitemOpen
  \bibfield  {author} {\bibinfo {author} {\bibfnamefont {M.~Y.}\ \bibnamefont
  {Niu}}, \bibinfo {author} {\bibfnamefont {I.~L.}\ \bibnamefont {Chuang}},\
  and\ \bibinfo {author} {\bibfnamefont {J.~H.}\ \bibnamefont {Shapiro}},\
  }\bibfield  {title} {\bibinfo {title} {Qudit-basis universal quantum
  computation using ${\ensuremath{\chi}}^{(2)}$ interactions},\ }\href
  {https://doi.org/10.1103/PhysRevLett.120.160502} {\bibfield  {journal}
  {\bibinfo  {journal} {Phys. Rev. Lett.}\ }\textbf {\bibinfo {volume} {120}},\
  \bibinfo {pages} {160502} (\bibinfo {year} {2018})}\BibitemShut {NoStop}%
\end{thebibliography}%

\end{document}